\documentclass[traditabstract,aas_macros]{aa}
\usepackage{times,graphics,graphicx,amssymb,paralist,natbib,threeparttable,bm,hyperref,caption,placeins,array,hyphenat,amsmath,multirow}
\usepackage{txfonts}
\providecommand{\adsurl}[1]{\href{#1}{ADS}} 
\providecommand{\url}[1]{\href{#1}{#1}} 

\PassOptionsToPackage{linktocpage}{hyperref}

\usepackage{txfonts}

\def\aap{A\&A}

\def\apjl{ApJ}

\def\apjs{ApJS}

\def\lesssim{\mathrel{\hbox{\rlap{\hbox{\lower4pt\hbox{$\sim$}}}\hbox{$<$}}}}
\def\gesssim{\mathrel{\hbox{\rlap{\hbox{\lower4pt\hbox{$\sim$}}}\hbox{$>$}}}}
\def\imagetop#1{\vtop{\null\hbox{#1}}}

\begin{document}  
\title{Abell~611 
\\ II.  \mbox{X-ray} and strong lensing analyses}
\author{
A.~Donnarumma\inst{1,2,3},
S.~Ettori\inst{2,3},
M.~Meneghetti\inst{2,3}, 
R.~Gavazzi\inst{4,5}, 
B.~Fort\inst{4,5},
L.~Moscardini\inst{1,3},\\ 
A.~Romano\inst{6,7},
L.~Fu\inst{8,10}, 
F.~Giordano\inst{7},
M.~Radovich\inst{8,9},  
R.~Maoli\inst{6},   
R.~Scaramella\inst{7}
and J.~Richard\inst{11}}

\offprints{A.~Donnarumma -- e-mail: annamaria.donnarumm2@unibo.it}

\institute{Dipartimento di Astronomia, Universit\`a di Bologna, via Ranzani 1,
40127 Bologna, Italy 
\and
 INAF-Osservatorio Astronomico di Bologna, via Ranzani 1, 40127 Bologna, Italy
 \and
 INFN, Sezione di Bologna, viale Berti Pichat 6/2,
40127 Bologna, Italy
\and
Institut d'Astrophysique de Paris, CNRS UMR 7095, 98bis Bd Arago, 75014 Paris, France
\and 
UPMC Universit\`e Paris 06,  UMR 7095, 75014 Paris, France
\and
 Dipartimento di Fisica, Universit\`a La Sapienza, Piazzale A.Moro 00185 Roma, Italy
 \and
 INAF-Osservatorio Astronomico di Roma, via Frascati 33, 00044 Monte Porzio Catone (Roma), Italy
 \and
 INAF-Osservatorio Astronomico di Napoli, via Moiariello 16, 80131 Napoli, Italy
\and 
INAF-Osservatorio Astronomico di  Padova, vicolo dell'Osservatorio 5, 35122 Padova, Italy
\and
Key Lab for Astrophysics, Shanghai Normal University, 100 Guilin Road, 200234, Shanghai, China
\and
Department of Physics, Institute for Computational Cosmology, Durham University, South Road, Durham DH1 3LE, United Kingdom}

\date{Accepted on October 17, 2010}

\abstract{
 We present  the results of our analyses of the \mbox{X-ray} emission and of the strong lensing systems in the galaxy cluster Abell~611 [$z=0.288$]. This cluster is   an optimal candidate for a comparison of the mass reconstructions obtained through \mbox{X-ray} and  lensing techniques, because of its very relaxed dynamical appearance and  its exceptional strong lensing system.
 We infer the \mbox{X-ray} mass estimate deriving  the density and temperature profile of the intra--cluster medium within the radius r$\simeq$700 kpc through a non-parametric approach,  taking advantage of the high spatial resolution of a \textit{Chandra} observation. Assuming that the cluster is in hydrostatic equilibrium and adopting a matter density profile,  we can recover the total mass distribution of Abell~611 via the \mbox{X-ray} data.   Moreover, we  derive the total  projected  mass  in the central regions of Abell~611 performing a parametric analysis of its strong lensing features through the publicly available analysis software \emph{Lenstool}. As a final step we compare the results obtained with both methods. We derive a good agreement between the \mbox{X-ray} and strong lensing total mass estimates in the central regions where the strong lensing constraints are present (i.e. within the radius r$\simeq$100 kpc), while  a  marginal disagreement is found between the two mass estimates when extrapolating the strong lensing results in the  outer spatial range. We suggest that  in this case the \mbox{X-ray/strong} lensing mass disagreement can be explained by an incorrect estimate of the relative contributions of the baryonic component and of the dark matter, caused by the intrinsic degeneracy between the different mass components in the strong lensing analysis.  We discuss the effect of some possible systematic errors that influence both mass estimates. We find a slight dependence of the measurements of the \mbox{X-ray} temperatures (and therefore of the \mbox{X-ray} total masses) on the background adopted in the spectral analysis, with total deviations on the value of $\rm M_{200}$ of the order of the 1$\sigma$ statistical error.
  The strong lensing  mass results are instead sensitive to the parameterisation of the galactic halo mass in the central regions, in particular to the modelling of the Brightest Cluster Galaxy (BCG) baryonic component, which induces a significant scatter in the strong lensing mass results.
}


\titlerunning{X-ray and strong lensing analyses  of Abell~611}
\authorrunning{A.~Donnarumma et al.}

\maketitle

\section{Introduction}
\label{sec:intro}
\begin{figure*}
\begin{center}
\includegraphics[width=0.9\textwidth]{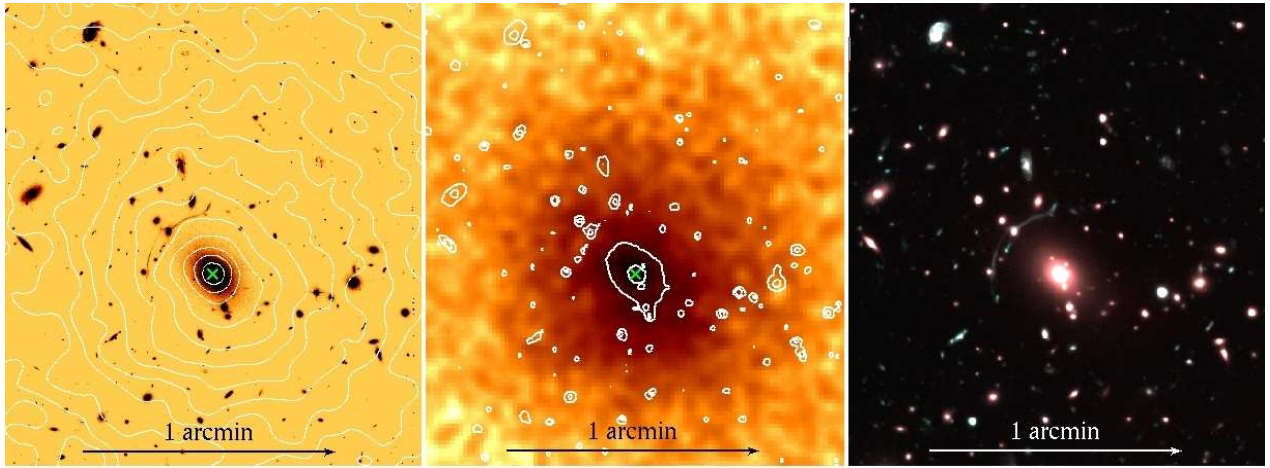} 
 \end{center}
 \caption[]{\textit{[Left panel]}  F606W \textit{ACS/HST} image of the cluster Abell~611 (see Table~\ref{tab:obs2} for the observation summary). The \mbox{X-ray} isophotes derived from the  \textit{Chandra} image (shown in the middle panel) are overlaid in white. The green cross indicates the \mbox{X-ray} centroid.  \textit{[Middle panel]}   The  \textit{Chandra} \mbox{X-ray} image of Abell~611 in the energy range [0.25-7.0] keV. The image was smoothed with a 3 pixel FWHM Gaussian filter, and is not exposure-corrected. The contours derived from the \textit{HST} image (shown in the left panel) are overlaid in white to facilitate the comparison between the images. Here again the green cross marks the \mbox{X-ray} centroid. \textit{[Right panel]}  Pseudo-colour composite image of Abell~611, obtained combining  $r$ and $g\!-\!\rm SLOAN$ images, taken with the Large Binocular Camera mounted at LBT. The three panels are WCS-aligned; the size of the field of view is $\simeq [1.6 \times 1.8]$ arcmin.}
 \label{fig:colour}
\end{figure*}

Testing the accuracy of the several approaches used to estimate  the  mass of galaxy clusters  is necessary to assess the reliability of methods involving  clusters as cosmological probes. For example, the comparison of the cluster baryon fractions $\rm f_{b}$  to the cosmic baryon fraction can provide a direct constraint on the mean mass density of the Universe, $\Omega_{\rm M}$ \citep{ettori2009,allen2008,ettori2003b}, while the evolution of the cluster mass function can tightly constrain  $\Omega_{\rm \Lambda}$ and the dark energy equation-of-state parameter $w$ \citep{vikhlinin2009a}. \\
Cluster mass profiles can be probed through several independent techniques, relying on different physical mechanisms and requiring different assumptions. The comparison of results obtained with different methods and the (dis)agreement between them provide a useful check, and can give additional insights on the inner structure of clusters. For example, cluster masses can be estimated through the  projected phase-space distribution and velocity dispersion of cluster galaxies \citep{biviano2009,diaferio2005}; the  dynamical studies, however, are affected by the  strong degeneracy between the mass density profile and the velocity dispersion anisotropy profile, and thus require strong assumptions on the form of the velocity dispersion tensor. \\
Measurements of the Sunyaev-Zel'dovich effect \citep[hereafter SZE; ][]{sunyaev1972}  towards  galaxy clusters yield a direct measurement of  the pressure distribution of the cluster gas: the total gravitational mass can thus be determined combining this information with  the intra--cluster medium (hereafter ICM) temperature, estimated for example through the \mbox{X-ray} spectral analysis \citep{grego2001}. Because it is independent of the cluster redshift, the SZE can be an extremely powerful tool to investigate the mass of \mbox{high--redshift} clusters \citep{hincks2009}, but  accurate measurements of the SZE are still a challenge because of intrinsic limitations and systematic biases  (faint signal extending over a large angular size  and subject to several sources of contamination).  \\
So far, the analysis of the cluster \mbox{X-ray} emission and of the gravitational lensing effect are among the most promising techniques to estimate galaxy cluster masses. \\ 
The \mbox{X-ray} mass estimate is less biased compared with lensing-derived masses by projection effects, because the \mbox{X-ray} emission is traced by the square of the gas density \citep{gavazzi2005,wu2000}.
 However, \mbox{X-ray} measurements of  cluster masses imply the assumption of hydrostatic equilibrium of the ICM with the dark matter potential and of spherical symmetry of the cluster mass distribution. Hence the total mass profile can be inferred from the radial profiles of gas temperature and  density. \\ 
 On the other hand, the gravitational lensing effect allows for the  determination of the projected surface mass density  of the lens, regardless of its dynamical state or the nature of the intervening  matter. This effect is determined by all massive structures along the line of sight, so lensing mass measurements are subject to foreground and background contaminations. Moreover, the lensing effect is sensitive to features of the mass distribution such as its ellipticity and asymmetry and to the presence of substructures \citep{meneghetti2007b}. \\The combination of \mbox{X-ray} and lensing results could verify and strengthen the  findings on the cluster physics and masses  (see for example \citealp{allen1998,ettori2003,bradac2008}); but so far several works in literature claimed a  significant disagreement between strong lensing and \mbox{X-ray} mass estimates,   (\citealp{wu1996,smail1997,ota2004,voigt2006,gitti2007,halkola2008}) which would prevent any joint analysis. Many convincing explanations, even if not conclusive, have been suggested. \\ 
It is therefore clear that a comparison between independent mass estimates for galaxy clusters can provide a vital check of the assumptions adopted in the cluster analysis, and will possibly give us additional insights into the underlying physics of these objects. In particular, the comparison between lensing and \mbox{X-ray} mass in galaxy clusters is also directly related to the cosmological parameters, since  their ratio  depends on the cosmic density parameter, $\Omega_{\rm M}$, and the cosmological constant,  $\Omega_{\rm \Lambda}$, for their different dependence on the angular diameter distances, which are smaller in an $\Omega_{\rm \Lambda}$ dominated universe, as pointed out by  \cite{wu2000}. \\
Targets for  a comparative analysis should be relaxed, unperturbed objects to avoid biases deriving from the incorrect assumption of hydrostatic equilibrium (X-ray analysis) or from the contamination of secondary mass clumps or unresolved substructures (lensing analysis). \\ 
An ideal candidate for this kind of analysis is Abell~611. It is a rich, cool-core cluster at  $z\!=\!0.288$ \citep{struble1999}; its \mbox{X-ray} emission peak is  well centred on the Brightest Cluster Galaxy  (hereafter BCG; see left and central panels of Fig.~\ref{fig:colour}), and the \mbox{X-ray} isophotes appear quite regular,  with a low degree of ellipticity. These characteristics are generally considered good indicators of a relaxed dynamical state. Relaxed objects are quite rare at this redshift, so a detailed study of the properties of Abell~611 is even more appealing.\\
In this work we will focus on the comparison between the \mbox{X-ray} and strong lensing mass estimates. A weak lensing analysis of large field of view data, obtained with the Large Binocular Camera (LBC) mounted at the Large Binocular Telescope (LBT), is presented in \cite{romano2010}  (hereafter Paper I). Here we present  new \mbox{X-ray} and strong lensing mass estimates, based on a non-parametric reconstruction of gas density and temperature profiles obtained through the analysis of a \textit{Chandra} observation,  and on a strong lensing reconstruction respectively,  which was performed with the \emph{Lenstool} analysis software  \citep{kneib1996,jullo2007}.
\\ 
This paper is organised as follows. In Sect.~\ref{sec:xray}  we will introduce our \mbox{X-ray} analysis, focusing on the data reduction and on the method applied to recover the total and gas mass profiles, and discuss some possible systematic errors associated with the \mbox{X-ray} analysis.  The strong lensing analysis is presented in Sect.~\ref{sec:sl},  where we will briefly discuss our main findings. We will compare the \mbox{X-ray} and strong lensing results in Sect.~\ref{sec:comparison}: a comparison with previous analyses can be found in Sect.~\ref{sec:prev}. Finally, we will summarise our results and draw our conclusions in Sect.~\ref{sec:concl}. \\
Throughout this work we will assume a flat $\Lambda$CDM cosmology: the matter density parameter will have the value $\Omega_{\rm m}$=0.3, the cosmological constant density parameter $\Omega_{\Lambda}$=0.7, and the Hubble constant will be $H_{0}=70 \ \rm  km\ s^{-1} Mpc^{-1}$. At the cluster redshift and for the assumed cosmological parameters, 1 arcsec is equivalent to 4.329 kpc. Unless otherwise stated, all uncertainties are referred to a $68\%$ confidence level. 
\section{X-ray analysis}
\label{sec:xray}
\subsection{X-ray data reduction}
\label{sec:reduc}
 The \mbox{X-ray} analysis of  Abell~611  returns several indicators of a  relaxed dynamical state,  such as the absence of  substructures in the \mbox{X-ray} emission (which appears isotropic over  elliptical isophotes), a well-defined cool-core, and a smooth surface brightness profile. The \mbox{X-ray} emission is centred on the BCG: the  distance between the  \mbox{X-ray}  centroid and the BCG centre is $\simeq 1$ arcsec (the  uncertainty on this measure is comparable to the smoothing scale applied to the \mbox{X-ray} image to determine the centroid, i.e. 3 pixels $\simeq 1.5$ arcsec). Moreover, Abell~611 appears to be a radio--quiet cluster, because the  Giant Metrewave Radio Telescope (GMRT) Radio Halo Survey \citep[][ see  Sect.~\ref{sec:prev}]{venturi2007,venturi2008} failed to detect any extended  radio emission associated with this cluster. Since giant radio halos are the most relevant examples of non-thermal activity in galaxy clusters \citep{brunetti2009}, the absence of a  radio halo supports the hypothesis of thermal pressure support, connected to the hydrostatic equilibrium assumption required in the \mbox{X-ray} analysis. \\
  This cluster can therefore be considered an ideal candidate to highlight any issue with the current analysis techniques: it fullfills the assumptions of thermal  pressure support and of spherical symmetry, so it is unlikely that potential biases in the results are related to the limiting assumptions underlying the \mbox{X-ray} analysis.
We performed our  \mbox{X-ray} analysis on the only \textit{Chandra}  data set available, retrieved from the public archive (see Table~\ref{tab:obs} for observation log). 
\begin{table}
\caption{Observation summary of the \textit{Chandra} exposure used for the \mbox{X-ray} analysis.}
\label{tab:obs}
\begin{center}
\begin{tabular}{l@{\@{\hspace{7pt}}}c@{\@{\hspace{8pt}}}c@{\@{\hspace{8pt}}}c@{\@{\hspace{8pt}}}c} \\
\hline\hline\noalign{\smallskip}
Obs.ID & Start Time & Total      & Net           & PI\\
       &            & Expos. [ks]  & Expos.[ks]  & name\\
\noalign{\smallskip}\hline\noalign{\smallskip}
 \vspace{3pt}3194   &  2001-11-03 18:43:58  & 36.6 & 30.9      & Allen \\
\hline		      
\end{tabular}
\end{center}
\end{table}
 We reduced the data using the \textit{CIAO} data analysis package -- version 4.0 -- and the calibration database CALDB 3.5.0\footnote{See the CIAO analysis guides for the data reduction:\\  cxc.harvard.edu/ciao/guides/ .}. We will summarise here briefly the reduction procedure. \\ The \textit{Chandra}  observation was  tele-metered in very faint mode and was performed through both  back-illuminated (BI) and front-illuminated (FI) chips. The cluster centre was imaged in the S3 BI chip.
  The level-1 event files  were reprocessed to apply the appropriate gain maps and calibration products and to reduce the ACIS quiescent background\footnote{For a complete discussion on this topic, see \\  cxc.harvard.edu/cal/Acis/Cal\_prods/vfbkgrnd/index.html .}. We used the \mbox{\texttt{acis\_process\_events}} tool to check for cosmic-ray background events and to correct for eventual spatial gain variations caused by charge transfer inefficiency to re-compute the event grades. We applied the standard event selection, including only the events flagged with grades $0,2,3,4$,  and filtered for the Good Time Intervals associated with the observation. The bright point sources were masked out, and a geometrical correction for the masked areas was applied; the point sources were identified using the script \texttt{vtpdetect}, and the result was then checked through visual inspection.  \\ 
A careful screening of the background light curve is necessary for a correct background subtraction \citep{markevitch2003}, to discard contaminating flare events, which could have a \mbox{non-negligible} influence on the  inferred cluster spectral properties.
We extracted the background light curve, applying a  time binning of about  1 ks, in the energy range [2.5--7] keV, which is the most effective band to individuate common flares for S3 chip.  We applied the script  \texttt{lc\_clean} to include only the time periods  within a factor 1.2 of the  mean  quiescent rate. We compared the S3 background light curve with the light curve extracted in the S1 chip using the energy range [2.5--6] keV, to check for  faint, long flares, which were not identified.  The S3 background light curve was examined using the \textit{ChIPS} facilities to identify and exclude further flaring events.  The net exposure time after the light curve screening is about 31 ks (see Table~\ref{tab:obs}).
\subsection{X-ray analysis}
\label{sec:x_a}
In order to derive a reliable cluster mass estimate through the \mbox{X-ray} analysis, a primary issue is to avoid any  bias in the temperature estimate, because cluster masses derived by assuming  hydrostatic equilibrium are dependent on the  temperature profile, as demonstrated by \citet{rasia2006}. 
For this purpose, a correct background modelling is crucial. 
The \mbox{X-ray} background is often estimated through the blank-sky background data sets provided by the ACIS calibration team. The blank-sky observations are reduced similarly to the source data sets to be compatible with the cluster event files; the former are then re-normalised by comparing the blank-sky count rate in a given energy range (generally $kT > 8.0$ keV, since in this band the \textit{Chandra} effective area is negligible) with the local background count rate. One of the advantages of this modus operandi is that the derived ARF and RESPONSE matrices will be consistent both for the source and  the background spectrum.   However, the background in the \mbox{X-ray}  soft band  can vary both in time and in space, so it is important to verify whether the  background  derived by the  blank-sky datasets is consistent with the ``real''  one. \\
For this purpose, we extracted a spectrum from the \emph{Chandra} observation in a region where the cluster emission is negligible: we compared the former to a spectrum derived in the same region of the blank-field data set.  We tried to fit the residuals in the [0.4--3] keV  band with a MEKAL model, without an absorption component and broadening the normalisation fitting range to negative values. 
This comparison shows some differences  between the two spectra. For this reason, we decided to proceed in the spectral fitting utilising both  a local background estimate and a blank-sky estimate.  The effect of the background choice in the mass estimate results will be discussed in Sect.~\ref{sec:sysx}.
\subsection{Spectral fitting}
\begin{figure*}
\begin{center}
 \begin{tabular}{cc}
 \imagetop{\includegraphics[width=0.46\textwidth]{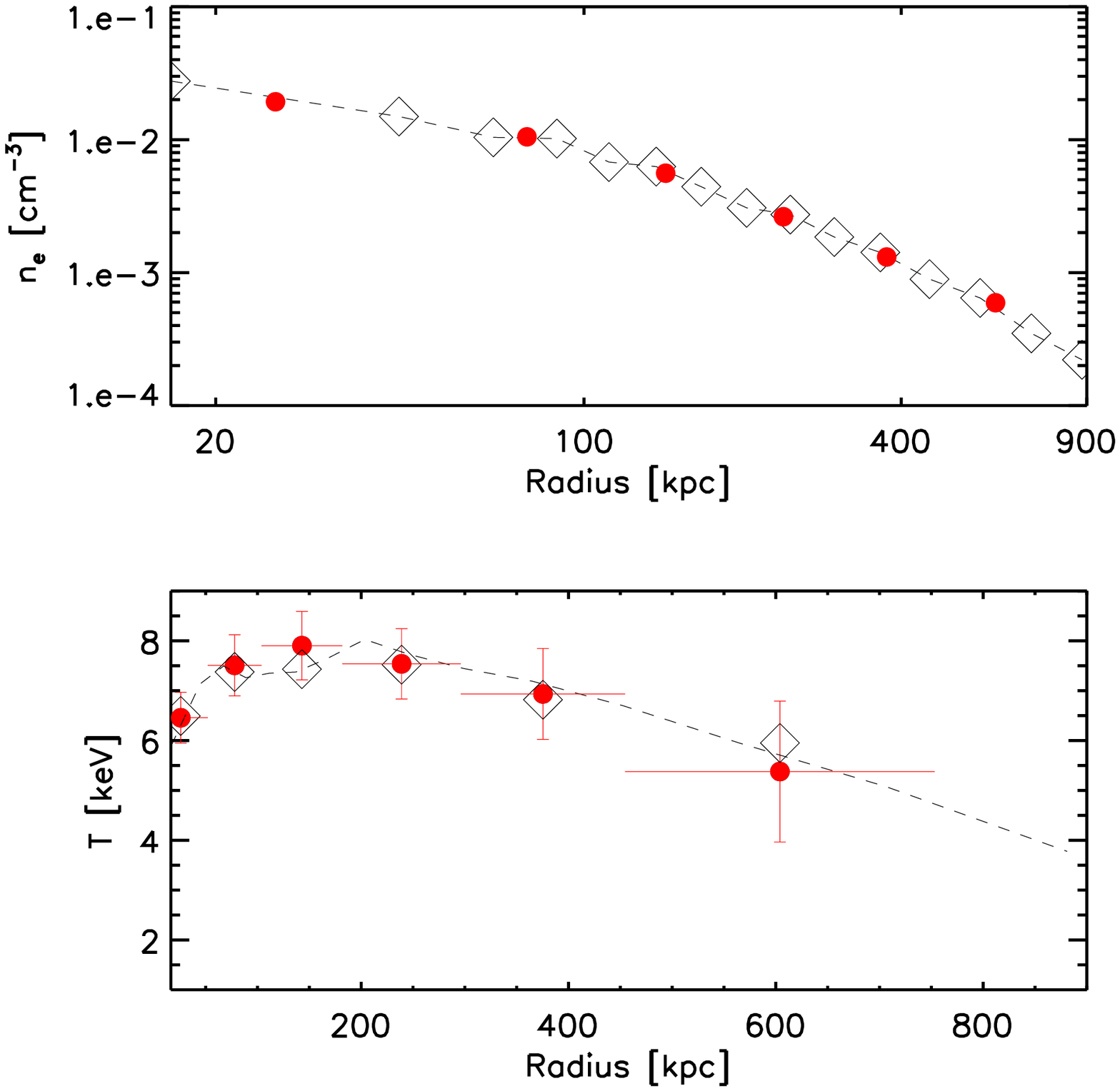}} & \imagetop{\includegraphics[width=0.44\textwidth]{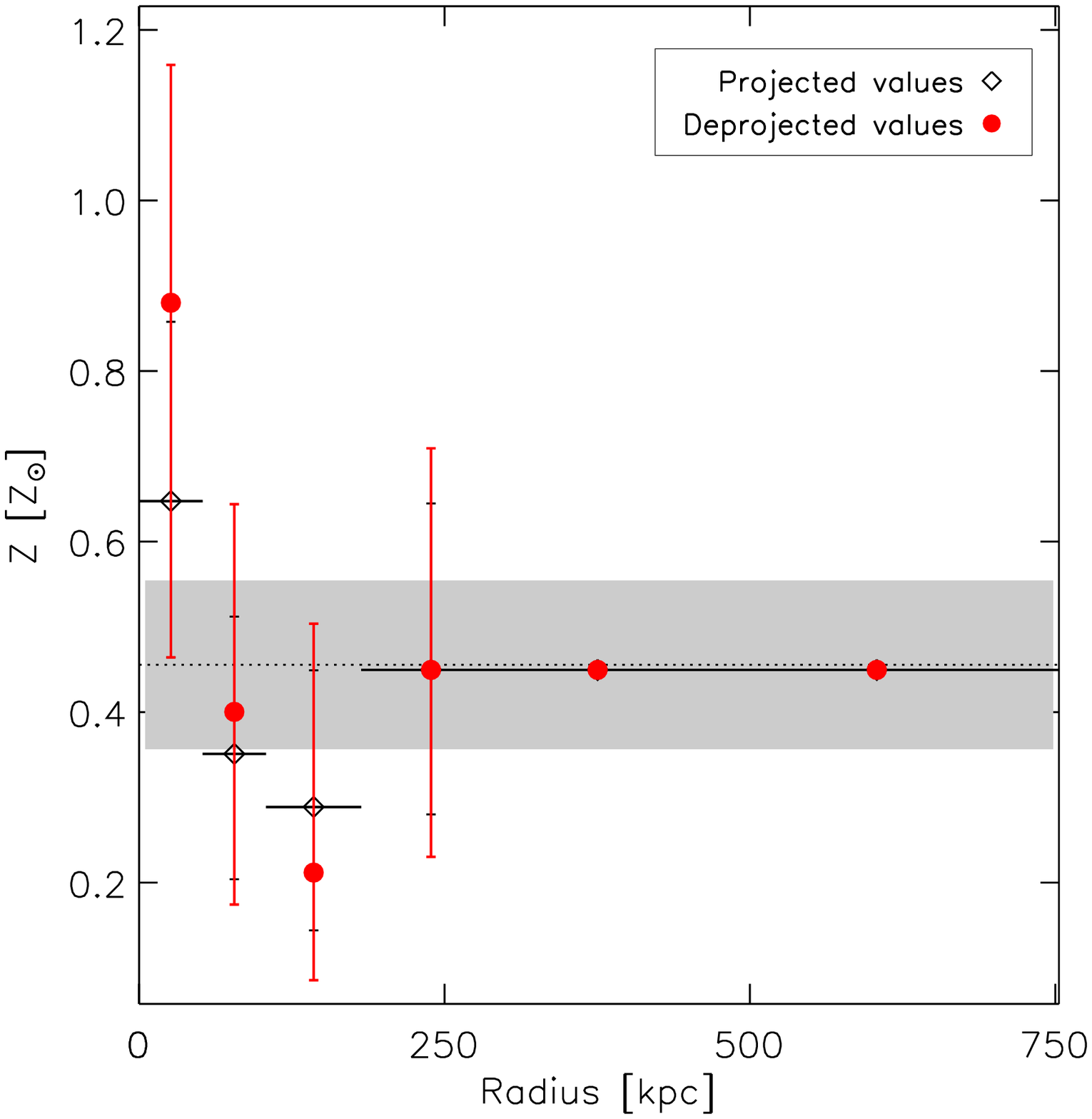}} \\
\end{tabular}
 \end{center}
\caption{\textit{[Top left panel]} Deprojected gas density profile. The red circles indicate the values determined by the normalisation of the thermal spectra, while diamonds represent the  values obtained combining both spectral and  surface brightness measurements: the dashed line shows the best-fit profile.  \textit{[Bottom left panel]}   ICM temperature profile. The figure shows the projected temperature values, directly inferred from the spectral analysis (red filled circles) and the  deprojected values, which were rebinned to the same binning of the spectral profile (diamonds). The  dashed line shows the deprojected best-fit profile.
\textit{[Right panel]}  The projected (black empty diamonds) and deprojected (red filled circles) metallicity profile for Abell~611 inferred from the \mbox{X-ray} analysis. The dotted line represents the mean metallicity value $Z=(0.45\pm 0.1)\ Z_{\odot}$, as determined through the fit of the spectrum extracted in the area with radius $R$= 2.9 arcmin, i.e. the cluster area covered by our spectral analysis. The grey region indicates the $1\sigma$ confidence level.}
 \label{fig:prof1}
\end{figure*}

  The spectra were extracted on concentric  annuli centred  on the \mbox{X-ray} surface brightness centroid, each one containing at least 2500 net source counts, to infer an estimate of the ICM metallicity. We selected six annuli with adequate S/N ratio, up to $R_{\rm spec}\!=\!2.9$ arcmin.  We used the \emph{CIAO} \texttt{specextract} tool  to extract the source and background spectra and to construct ancillary-response and response matrices. The spectra were fitted  in the $0.6 - 7.0$ keV range except for the last two annuli, in which, due to the higher background level, we restricted the analysis to the range $0.6 - 5.0$ keV.
   We performed the spectral analysis using the \emph{XSPEC} software package  \citep{arnaud1996}. We adopted an optically-thin plasma emission model (the \texttt{mekal} model; \citealp{mewe1985,kaastra1993}) with an absorption component (the T\"ubingen--Boulder model \texttt{tbabs}; \citealp{wilms2000}).  \\
We fixed the Galactic absorption to the value inferred from radio HI maps in \cite{dickey1990}, i.e. $4.99\times 10^{20}$ cm$^{-2}$.  The free parameters in the spectral fitting model are the temperature, the metallicity and the normalisation of the thermal spectrum. \\
The metal abundance profile presents a central peak with a quite scattered trend (see the right panel of  Fig.~\ref{fig:prof1}). The metallicity profile is not directly involved in the \mbox{X-ray} mass reconstruction: but we still verified whether a bias in its measurement can influence the derived mass estimate.  The results of this check will be presented in Sect. ~\ref{sec:sysx}.
\subsection{X-ray mass profile}
\begin{figure*}
\begin{center}
\includegraphics[width=0.5\textwidth]{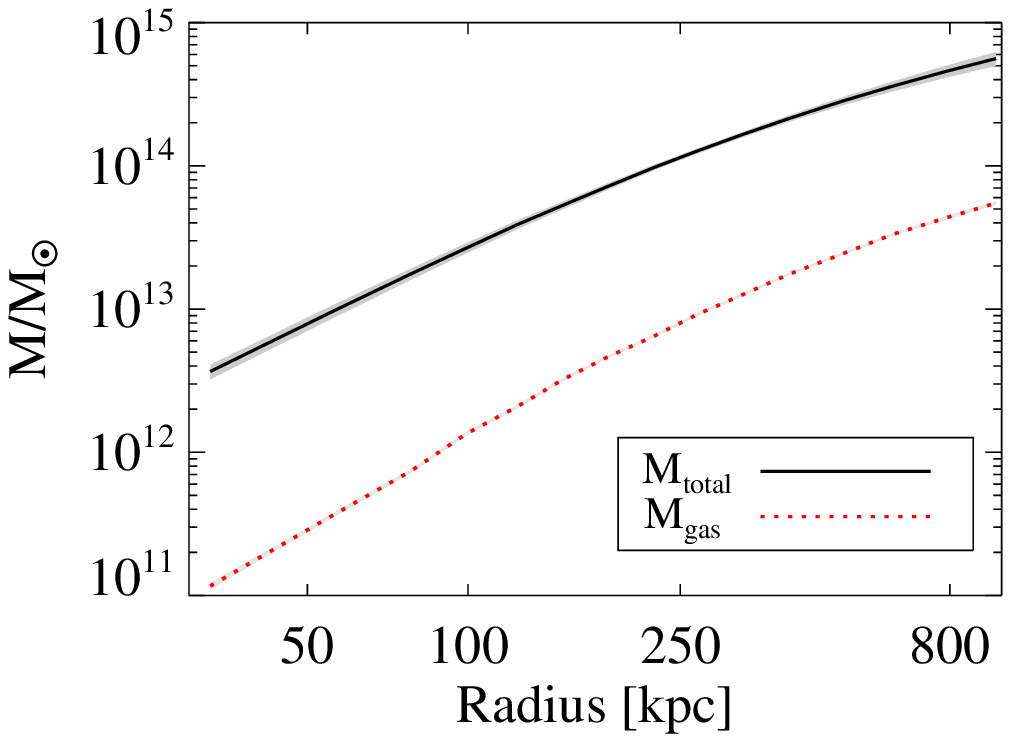}\hspace{-0.6cm}\includegraphics[width=0.5\textwidth]{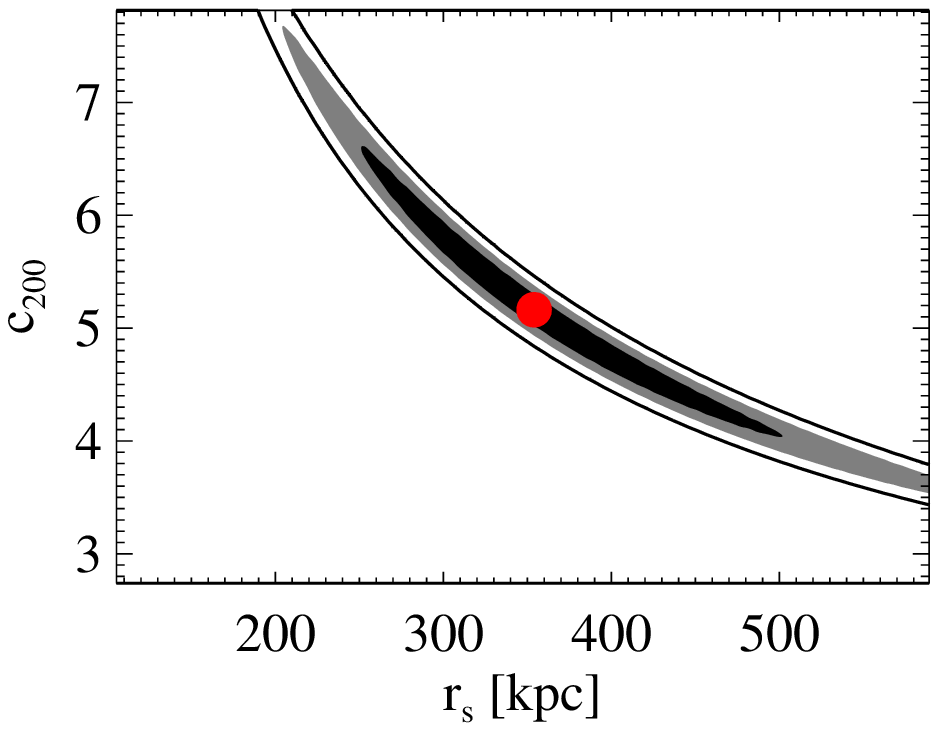}
 \end{center}
\caption[]{\emph{[Left panel]} Mass profiles for Abell~611 as determined through the \mbox{X-ray} analysis. The black solid line  and the red dashed line indicate the total mass and the gas mass profile, respectively. \emph{[Right panel]} Confidence levels determined through the \mbox{X-ray} analysis for the NFW profile parameters $r_{\rm s}$ and $c_{200}$.  The red circle indicates the best-fit values, while the contours represent the $1\sigma,2\sigma$, and $3\sigma$ confidence  levels.}
 \label{fig:res_x}
\end{figure*}

The gas temperature and density profiles are recovered  without any parameterisation of the ICM properties, but relying on the assumptions of spherical symmetry and  hydrostatic equilibrium and on the choice of a  model for the total mass. \\
 The method that was applied is briefly resumed in the following; the reader can refer to \citet*{ettori2002,morandi2007a,morandi2007b} for full details.\\
In this section, the subscript ``ring'' will indicate a  projected quantity measured in concentric rings (equivalently, spectral bins), whereas   the subscript ``shell'' will point to a deprojected quantity measured in  volume shells.
The projected gas temperature $T_{\rm ring}$, metal abundance $Z_{\rm ring}$ and the normalisation $K_{\rm ring}$ of the model for each annulus are derived  through the spectral analysis. In particular, the normalisation of the thermal spectrum provides an estimate for the emission integral $EI = \int n_{\rm e} n_{\rm p} {\rm d}V \simeq   0.82 \int n^{2}_{\rm e}\, {\rm d}V$. \\
To deproject the former quantities, we need to derive the  intersection of each volume shell  with the spectral annuli, which can be calculated through an upper triangular matrix  $\mathcal{V}$, whose last pivot represents the outermost annulus if we assume spherical symmetry  \citep*{kriss1983,buote2000}.\\
To calculate the deprojected density profile, we  combined the spectral information and  the cluster \mbox{X-ray} surface brightness (SB) profile, composed by 15 bins with 1000 counts each.  The  SB provides an estimate of the  volume-counts emissivity $F_{\rm bin} \propto n_{\rm e,\rm bin}^{2}T_{\rm bin}^{1/2}/D^{2}_{\rm L}$, where $D_{\rm L}$ is the luminosity distance of the cluster. By comparing this observed profile  with the values predicted  by a thermal plasma model, with  temperature and metallicity equal to the measured spectral quantities and taking into account the absorption effect, we can solve for the electron density and obtain  a gas density profile better resolved than a ``spectral--only'' one (see the top panel of Fig.~\ref{fig:prof1}).  
\\ The total gravitating mass was then constrained as follows: the observed deprojected temperature profile in volume shells $T_{\rm shell}$ was obtained through
\begin{equation}  
\epsilon T_{\rm shell} = (\mathcal{V}^T)^{-1} \# (L_{\rm ring} T_{\rm ring}), 
\end{equation}  
where 
\begin{equation}  
 \epsilon = (\mathcal{V}^T)^{-1} \# L_{\rm ring};
\end{equation}  
\noindent here the symbol $\#$ indicates a matrix product, $\mathcal{V}$ is the volume matrix and  $L_{\rm ring}$ is the cluster luminosity measured in concentric rings.
We compare it to the predicted temperature profile  $T_{\rm model}$, obtained by inverting the equation of hydrostatic equilibrium between the dark matter  and the intracluster plasma, i.e.:
\begin{eqnarray}
-G \mu m_{\rm p} \frac{n_{\rm e} M_{\rm tot, model}(<r)}{r^2} =
\frac{{\rm d}\left(n_{\rm e} \times kT_{\rm model}\right)}{{\rm d}r},
\label{eq:mtot}
\end{eqnarray}
where $n_{\rm e}$ is the deprojected electron density. Our best-fit mass model was determined  by comparing the observed temperature profile $T_{\,\rm obs}$ with the predicted profile $T_{\,\rm model}$, which depends on the mass model parameters:  accordingly we minimised the quantity:
\begin{equation} 
\chi^2 = \sum_{\,\rm \,rings}\,\left(
\frac{T_{\,\rm obs} - T_{\,\rm model}}{\sigma_{\,\rm obs}} \right)^2,
\label{eq:chi2x}
\end{equation} 
where the $T_{\rm model}$ profile was rebinned to  match the observed temperature profile.
In the bottom panel of Fig.~\ref{fig:prof1} we show the projected temperature profile as determined through the spectral analysis, the deprojected best-fit profile and the deprojected profile, rebinned to the spectral intervals. \\
We assumed the NFW profile as the total mass functional; this profile can be expressed as:
\begin{equation}
\rho(r)=\frac{\rho_s}{(r/r_s)\left(1+r/r_s\right)^2}\;,
\label{eq:NFW}
\end{equation}
\noindent where  $\rho_s$ is the characteristic density and $r_{\rm s}$ is the scale radius. 
We will  describe the NFW profile through the scale radius and  the concentration parameter $c_{200}$; the former can be defined  as
 \begin{equation}
 c\equiv \frac{R_{200}}{r_{\rm s}}\;,
\end{equation}
 where $ R_{200}$ is the radius within which the mean  matter density of the halo is 200 times the critical density of the Universe.
 Assuming a mass model, the  best-fitting values of its parameters are determined by minimising the quantity defined in Eq.~\ref{eq:chi2x}.       
 The NFW profile parameters were optimised in the ranges $0.5\!\leqslant\!c_{200}\!\leqslant 20$ and $10\ {\rm kpc}\leqslant\! r_{\rm s}\!\leqslant\!976\ {\rm kpc}$, where the scale radius upper limit is the outer radius of the surface brightness profile.
The  gas and total mass profiles are shown in Fig.~\ref{fig:res_x}; the inferred mass model parameters are listed in Table~\ref{tab:tab_x1}. 
\begin{table*}[!htb]
\caption[]{Best-fit values for the total mass parameters obtained through the \mbox{X-ray} analysis.}
\begin{center}
\begin{tabular}{lcccc}
 \multicolumn{5}{c}{\textsc{Results of the X-ray  analysis of Abell~611}} \\
\noalign{\smallskip}\hline \noalign{\smallskip}
 & $\rm r_{s}$ & $\rm c_{200}$ & $\rm M_{200}$ & $\chi^{2}_{\rm red}$[d.o.f.]\\
 & [kpc] & & [$10^{14}\rm M_{\odot}$] &\\ 
\noalign{\smallskip}\hline
\hline\noalign{\smallskip}
(1)~~Blank-sky background  &  $350.3\pm 79.6$ & $5.18\pm 0.84$ & ~~$9.32\pm 1.39$ & $0.81~~[4]$ \\ 
(2)~~Local background  & $420.0\pm99.5$&$   4.67\pm0.80$ &   $11.11\pm2.06$ & $0.46~~[4]$ \\
(3)~~Fixed metallicity  & $367.2\pm98.6$ & $   5.13\pm0.87$ & $10.61\pm1.96$ & $0.33~~[4]$\\
\noalign{\smallskip}\hline
\end{tabular}

\end{center}
\label{tab:tab_x1}
\tablefoot{ 
(1) the blank-sky dataset is adopted to estimate the \mbox{X-ray} background; \\
(2) the \mbox{X-ray} background is derived from peripheral regions of the source observation; \\
(3) the background is estimated locally fixing the gas metallicity to $Z=0.3\ Z_{\odot}$. \\
The columns indicate the best-fit values obtained for the NFW profile parameters 
(the scale radius $r_{\rm s}$ and the concentration $c_{200}$), 
the estimate of the total mass within $R_{200}$ (here listed as  $M_{200}$) 
and the reduced $\chi^{2}$, referred to a fit with four degrees of freedom. }
\end{table*}
 The  confidence levels for the NFW  parameters are shown in the right panel of Fig.~\ref{fig:res_x}. 
\subsection{Systematic effects in the \mbox{X-ray} mass  profile}
 \label{sec:sysx}
\begin{figure}
\begin{center}
\includegraphics[width=0.5\textwidth]{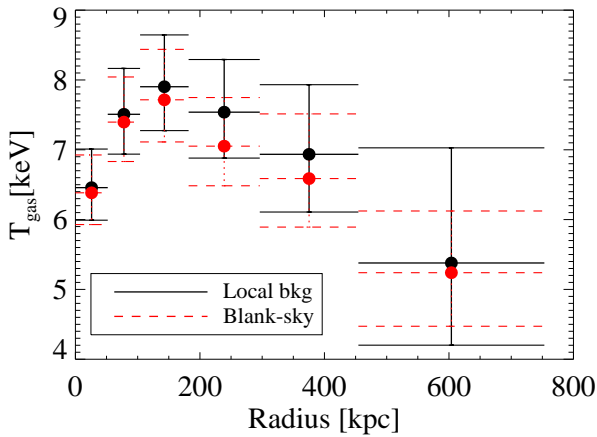} 
\end{center}
\caption[]{Comparison between the temperature profile determined by estimating the \mbox{X-ray} background  through the blank-sky dataset (red dashed line) or using a local background component (black solid line).}
\label{fig:temp_comp}
\end{figure}

 In this section we will briefly  assess if our  \mbox{X-ray} mass estimate of Abell~611 can be affected by some analysis choices. \\ 
 The choice of the \mbox{X-ray} background estimate, as discussed in Sect.~\ref{sec:x_a}, could have a \mbox{non-negligible} impact on the derived temperature profile, and consequently on the total mass measurement.
  Moreover,  we  verified whether considering the ICM metallicity as a free parameter in the spectral analysis can impact the  \mbox{X-ray}  mass estimate of Abell~611, since (as shown in Fig.~\ref{fig:prof1}) its inferred metallicity profile  is quite irregular. We thus re-ran the spectral analysis by fixing the metallicity $Z$ to the typical value of $0.3Z_{\odot}$ \citep[see e.g. ][]{balestra2007}; we then derived a new \mbox{X-ray}  mass estimate adopting the spectral profiles obtained under the assumption of a constant metallicity. \\
   Therefore we derived the mass of Abell~611 through the \mbox{X-ray} analysis:
   \begin{enumerate}
 \item  by  estimating the \mbox{X-ray} background through the spectra extracted from the blank-sky dataset (opportunely reprocessed and re-normalised to be consistent with the source observations) in the same image regions used for the source spectra (case 1); 
 \item by estimating  the  \mbox{X-ray} background  from peripheral regions of the S3 chip in the source  dataset (case 2);
  \item  as in the previous case, but by fixing the metallicity $Z$ to the value $0.3Z_{\odot}$ (case 3).
 \end{enumerate}  
  The NFW parameters derived from the \mbox{X-ray} analysis in the three aforementioned cases are listed in  Table~\ref{tab:tab_x1}. \\
 The results obtained in the first two  cases mutually agree  within the 1$\sigma$ range, even if the temperature profile derived in the latter case is systematically higher (see Fig.~\ref{fig:temp_comp}). 
   Moreover, the best-fit parameters obtained in the third case (i.e. fixing the metallicity to $0.3Z_{\odot}$: see case 3  in Table~\ref{tab:tab_x1}) agree with the values obtained by including the metallicity as a free  parameter in the  fitting: the choice of estimating or fixing the metallicity in the spectral analysis has a negligible impact on the derived cluster mass. \\
\section{Strong lensing analysis}
 \label{sec:sl}
\subsection{Imaging data}
Our strong lensing analysis of Abell~611  benefits from the high-resolution imaging data obtained with the Hubble Space Telescope (\textit{HST}), which allowed an unambiguous identification of several lensed systems.
Abell~611 was observed with the Advanced Camera for Surveys (\textit{ACS}) mounted on  \textit{HST}. We retrieved the  \textit{ACS} association data of Abell~611 from the Multimission Archive at STScI (MAST)\footnote{The  url of the MAST is archive.stsci.edu.}; the details of the \textit{HST} observation of Abell~611 are summarised in Table~\ref{tab:obs2}. \\The  \textit{HST} archival data for Abell~611 are already processed following the  \textit{ACS} pipeline and combined using the \texttt{multidrizzle} software \citep{koe2002}, with a final pixel scale of 0.05 arcsec; consequently no additional basic data reduction is necessary. But we used the softwares \textit{SExtractor} \citep{bertin1996}, \textit{Scamp} \citep{bertin2006} and \textit{Swarp}\footnote{The software \textit{SExtractor}, \textit{Scamp} and \textit{Swarp} can be found at the url: astromatic.iap.fr/software/sextractor, astromatic.iap.fr/software/scamp and astromatic.iap.fr/software/swarp, respectively.} \citep{bertin2002}, developed at the TERAPIX   data reduction centre to derive the object catalogue for the Abell~611 \textit{HST} imaging data, re-compute an astrometric solution and finally resample the image to derive a more accurate astrometric matching and to equalise the background level. The WCS coordinates reported in the present work are  given with regard to the  \textit{HST} data and refer to  the astrometric solution that we re-derived.
\begin{table}
\caption{Observation summary of the \textit{ACS/HST} image used for the strong lensing analysis.}
\label{tab:obs2}
\begin{center}
\begin{tabular}{lll} 
\hline \hline\noalign{\smallskip}
Data-set ID &~~~~&J8D102010 \\
Start Time &~~~~&002-12-03 21:07:28 \\
Exposure [ks]   &~~~~&   2.16\\
Instrument &~~~~& ACS/WFC\\
Filter &~~~~& F606W \\
Proposal  ID    &~~~~&  9270\\
PI &~~~~& Allen\\
\hline	     
\end{tabular}
\end{center}
\end{table}
  Abell~611  was also observed during the Science Demonstration Time for the Blue Channel of the Large Binocular Camera (LBC), mounted at
the Large Binocular Telescope (LBT) (located at the Mt. Graham International Observatory, Arizona), in  good seeing conditions (FWHM of the seeing disc $\sim$ 0.6 arcsec).  The observations were made with  the $g$-SLOAN and $r$ filters; the total exposure times of the two coadded images are 3.6 ks and 0.9 ks, respectively.  The LBT data were reduced according to the data reduction pipeline developed at the INAF/OAR centre\footnote{See the Large Binocular Camera  website hosted by  the OAR-Osservatorio Astronomico di Roma: lbc.mporzio.astro.it .}; the astrometric solution was derived with the ASTROMC software \citep{radovich2008}. The final  plate scale of the LBC images is 0.225 arcsec/pixel. Further details on the LBC/LBT data and on their  reduction can be found in Paper I  and in \cite{giallongo2008}; a comparison with the weak lensing mass estimate obtained in Paper I  is presented in Sect.~\ref{sec:prev}.  The pseudo-colour image of Abell~611, obtained by combining the two LBT observations, is shown in Fig.~\ref{fig:colour}.  We relied on  the $g$-SLOAN and $r$  LBT observations  both  to verify whether the candidate conjugated images have similar colours, as it should be because of the intrinsic surface brightness conservation in gravitational lensing, and to select the cluster galaxies, in order to account for their mass in the lens modelling (see  $\S \ $~\ref{sec:gal}).
\subsection{Strong lensing system}
\label{sec:slsys}
\begin{figure}[!thb]
\begin{center}
\includegraphics[width=0.48\textwidth]{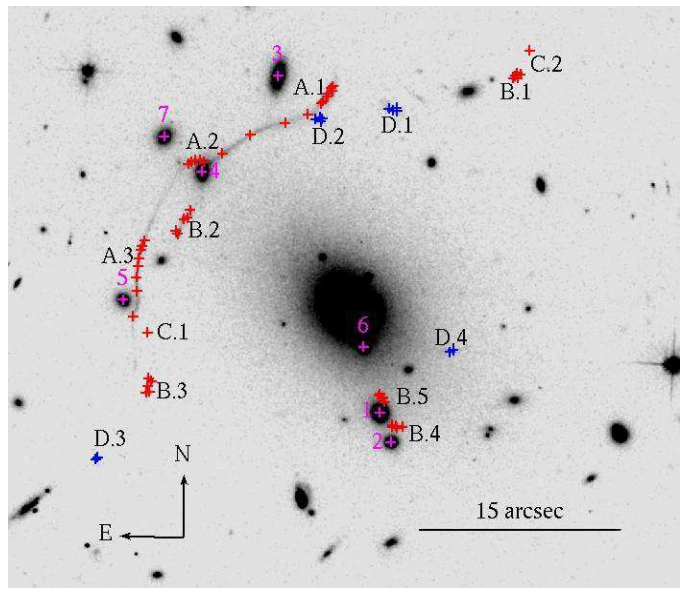}  
\end{center}
 \caption[]{Overview of the strong lensing system of Abell~611. The knots that we used as constraints are marked as red  crosses on the F606W \textit{ACS/HST} image; enlarged images  are shown in Fig.~\ref{fig:snap}. The magenta crosses indicate the perturber galaxies that were  optimised singularly (see Case 6 in Table~\ref{tab:tab_l1}). The blue crosses correspond to a likely lensed system, identified by R09, which  is not yet confirmed spectroscopically. We discuss our selection of the input images and its effect on the results in Sect.~\ref{sec:slsys} and Sect.~\ref{sec:slres}. The coordinates of all the systems are listed in Table~\ref{tab:input}. The size of the field of view is $\simeq 50 \times 42.5$  arcsec; the image was smoothed with a 2 pixel FWHM Gaussian filter (1 pixel = 0.05 arcsec).}
\label{fig:sysgen}
\end{figure}

Abell~611 exhibits  an outstanding strong lensing system. The most evident lensed feature is the giant tangential arc located at about 15.5 arcsec from the BCG centre (see Fig.~\ref{fig:sysgen}). 
The  curvature of the arc significantly changes in the proximity of  some cluster galaxies, which locally perturb the critical lines and possibly the magnification and the multiplicity of the lensed images \citep{keeton2003,amara2006,meneghetti2007a,peirani2008,puchwein2009}. \\
\cite{richard2009} (hereafter R09) obtained the optical spectra of three surface brightness spots along the tangential arc, deriving a redshift estimate of $z = 0.908\pm0.005$ for all of them.
The arc appears to be formed by three merging images (A.1, A.2 and A.3 in Fig.~\ref{fig:sysgen}), as in a typical cusp configuration. \\
 Another confirmed  lensed system  is a fold  system (hereafter called system  B). \cite{richard2009}  reported the spectroscopic redshifts obtained for images  B.2 and B.3  in Fig.~\ref{fig:sysgen}, confirming that  both of them are lensed images of a source at redshift  $z = 2.06\pm0.02$. The  images labelled B.1 and B.4/B.5  can be safely considered as their counter-images. More specifically, the image B.5 appears to be an additional image produced by  the position of the source B, lying totally or partially inside the caustic of the nearby  galaxy (galaxy n.1 in Table~\ref{tab:galpar}). 
 The g-r colours of all images of system B are similar (see Fig.~\ref{fig:colour}).\\
  A third lensed system was serendipitously identified by R09.  They were able to derive the redshift  $z_{\rm C.1}=2.59\pm0.01$. of image C.1 (see Fig.~\ref{fig:sysgen}) thanks to its strong Lyman-alpha emission. 
  The image C.2 (see  Fig.~\ref{fig:sysgen}) was considered by R09 to be the counter-image of C.1. We agree with their identification and included both images as constraints. \\
  Finally, the images labelled D.1 to D.4 in Fig.~\ref{fig:sysgen} were identified by R09 and by \cite{newman2009} (hereafter N09; see  Sect.~\ref{sec:prev} for more details about their recent study of Abell~611) as a lensed system, from the  similarity of the morphologies of images D.1 and D.2. However, this system is not yet confirmed spectroscopically, so we decided not to include it in the strong lensing analysis, but we verified if and how the best-fitting results would be modified by considering it as an additional constraint (see Sect.~\ref{sec:slres}). 
\subsection{Strong lensing modelling}
\label{sec:slan}
We performed a parametric reconstruction of the mass distribution of Abell~611 with the \emph{Lenstool}  analysis software\footnote{The \emph{Lenstool} software package  is available at  www.oamp.fr/cosmology/lenstool.} \citep{kneib1996,jullo2007}.  \\
 Our lens model consists of the superposition of a smooth cluster-scale potential   and of galaxy--scale halos, associated with  \mbox{early-type}  cluster galaxies.  The details of the cluster scale halo modelling will be discussed in $\S \ $~\ref{sec:clus}, whereas we will discuss the  galaxy-scale component  in $\S \ $~\ref{sec:gal}; finally the BCG modelling will be  discussed separately in $\S \ $~\ref{sec:bcg}.  \\
 We constrained the lens model for Abell~611  by identifying in  each image system, consisting of the multiple images of the same source, some elliptical regions or knots,  which are likely to be multiple images of the same source area.  The knots were identified through  visual inspection of the \textit{ACS/HST} observation by selecting the local peaks in the surface brightness, which are detectable in the strong lensing  images of Abell~611. The surface brightness spots were then grouped into sets of conjugated  knots by evaluating characteristics such as their relative positions  with respect to the location of the critical lines and  their brightness ratios. \\ 
  We assumed an astrometric positional error  of $\sigma_{\rm pos}\! =\! 0.2$ arcsec. The adopted error affects the $\chi^2$ estimator  computed in the source plane by \textit{Lenstool}: 
\begin{equation}
 \label{eq:chisl}
\chi^2_{S_i} = \sum_{j=1}^{n_i} 
	\left( \frac{x^j_\mathrm{S}-< x^j_\mathrm{S}>}
               {\mu_j^{-1}\sigma_{\rm pos}}\right)^2,
\end{equation}
 \noindent where $x^j_\mathrm{S}$ is the position of the source corresponding to  the
 observed image $j$, $<x^j_\mathrm{S}>$  the barycenter
 of the $n_i$ source positions (corresponding to the $n_i$ conjugated images or knots of the
image system $i$), and $\mu_j$  the
 magnification for the lensed image $j$;  the final $\chi^2$ is then obtained by combining the individual $\chi^2_{S_i}$ computed for each image system. \\
   Between the several optimisation methods available in  \textit{Lenstool} we adopted the Bayesian optimisation method, with a convergence speed parameter equal to 0.4; this method, although slower than the other optimisation options, is less sensitive to local minima in the parameter space.
\subsection{Cluster--scale component}
\label{sec:clus}
We modelled the cluster scale component assuming an elliptical NFW profile as mass functional.
The initial values for the centre, ellipticity $e$ and position angle
$\theta$ of the cluster halo were inferred from the BCG luminosity
profile. The centre was allowed to vary $\pm5$ arsec, while the
ellipticity and the position angle were optimised in the ranges $0.01 \leqslant e \leqslant 0.4$  and $100.0\ \rm deg\leqslant \theta_{\rm DM} \leqslant 140.0\ \rm deg$, respectively. The   optimisation range of the cluster position angle is asymmetric in order to take into account  the orientation of the X-ray isophotes.
 The optimisation ranges for the NFW profile parameters were  $c_{200} \in [0.5 -14.0]$ and  $r_{\rm s} \in [50.0 - 800.0]$ kpc.\\
\subsection{Galaxy--scale components}
\label{sec:gal}
\subsubsection{Selection of cluster galaxies}
\begin{figure}[!htb]
\begin{center}
\includegraphics[angle=-90,width=0.48\textwidth]{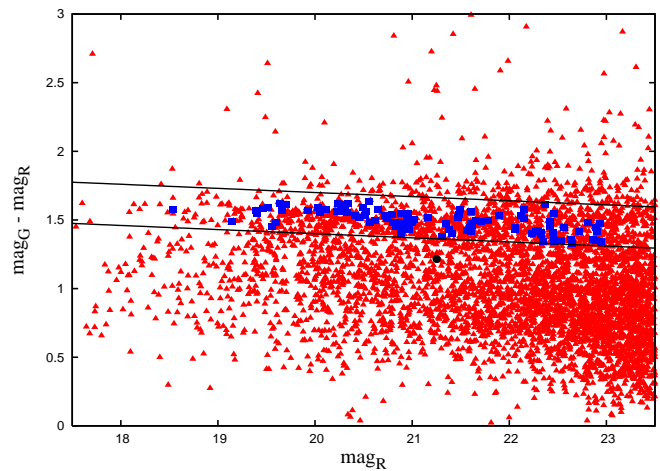} 
\end{center} 
 \caption[]{Colour-magnitude diagram of Abell~611. The black  lines define the cluster red-sequence, while the blue filled squares indicate the red sequence galaxies  lying within 100 arcsec from the BCG centre, which were included in the lens modelling. The black circle indicates the galaxy located between the lensed images B.1 and C.1 (see Fig.~\ref{fig:sysgen}), which was included in the lens modelling for its position; the magnitude estimate derived for this galaxy  from the LBT data is contaminated by the flux contribution of the nearby lensed feature (see text for details). 
   }
\label{fig:reds}
\end{figure}

Several authors demonstrated that the inclusion of galaxy-scale halos in the lens modelling (in particular the galaxies whose projected position is within the region probed by strong lensing) is necessary to account for their effect on the lensing cross section and  on the appearance of the lensed images that fall close to the perturbers \citep{natarajan2007,amara2006,meneghetti2003b,keeton2003,bradac2002}.
Although gravitational lensing is sensitive to all massive structures distributed along the line of sight, the main galactic mass budget is constituted of the massive  \mbox{early-type}  cluster galaxies; we will thus consider only this galactic population in our lens modelling.\\
The cluster members were selected through the characteristic cluster red-sequence, which allows us to identify the red  \mbox{early-type}  galaxies \citep{bell2004} and  which is clearly detected in the colour-magnitude diagram   (see Fig.~\ref{fig:reds}).
  We will assume in the modelling that all the red-sequence galaxies belong to the cluster, and thus lie at its same redshift. The diagram was derived by extracting the object catalogues from the two LBT observations, using the \textit{SExtractor} software in single and dual-image mode.\\
 Even if the total exposure time of the red  image is 0.9 ks, which is one fourth of the $g$-Sloan observation exposure time, the former filter allows an easier deblending of the redder cluster galaxies from the bluer lensed features.  For this reason we used the $r$ filter image for the detection of the sources in the dual-image mode. Moreover, even if the higher spatial resolution of \textit{ACS} data would guarantee a more accurate deblending of contiguous sources, we decided  to use the LBC data,  which cover a wider area, to derive the cluster galaxy catalogue. \\
 Only galaxies located within 100 arcsec ($\simeq$ 433 kpc) from the BCG centre  and whose magnitude $m_R$ is lower than 23 were included in the lens modelling; a total of 97 galaxies were selected adopting these criteria. We added to the catalogue the galaxy located between system B.2 and B.3 in Fig.~\ref{fig:sysgen}  that could affect the lensed images due to its position. The colour-magnitude diagram, the  cluster red-sequence and the galaxies included in the lens modelling are shown in  Fig.~\ref{fig:reds}. \\
\subsubsection{Estimate of cluster galaxy parameters}
We used  the two-dimensional fitting algorithm \textit{GALFIT}\footnote{The \textit{GALFIT} software is publicly available at the url users.obs.carnegiescience.edu/peng/work/galfit/galfit.html .} \citep{peng2002} to model and subtract the BCG as well as seven galaxies located within $\simeq 17$ arcsec from the BCG centre and with magnitude in the \textit{ACS} F606W filter lower than 22 for a better identification of the multiple lensed images and to extract the BCG surface brightness profile. All galaxies were fitted simultaneously; the lensed features and the point-like sources were masked out.  We modelled the galaxy light profile adopting both a S\'ersic and a de~Vaucouleurs profile. The S\'ersic profile \citep{sersic1968} has the functional form
\begin {equation}
\Sigma(r)=\Sigma_e \exp{\left[-\kappa \left(\left({\frac{r}{r_e}}\right)^{1/n} - 1\right)\right]},
\label{eq:sersic}
\end {equation}
\noindent where $\Sigma_e$ is the pixel surface brightness at the effective radius
$r_e$, the latter defined as the radius that encloses half of the total flux, and $n$  is the S\'ersic index. The  de~Vaucouleurs profile \citep{devauc1948} has the same analytical form as the  S\'ersic profile,  but the index has the fixed value of 4.\\
The galaxy models were convolved with the \textit{HST} PSF image created using the TinyTim software \footnote{The TinyTim software is available at the url  www.stsci.edu/software/tinytim/tinytim.html .} \citep{krist1993}. The reduced $\chi^2$ of the best-fitting models obtained adopting either the S\'ersic or the de~Vaucouleurs profile  are similar, $\chi^2_{S}=3.8$ and $\chi^2_{dV} = 3.9$, respectively. However, the galaxy-subtracted image derived fitting the galaxy light profile with a  de~Vaucouleurs model shows larger residuals in the central area for some of the fitted galaxies, in particular for the galaxy labelled 1 in Table~\ref{tab:galpar}. For this reason we chose a  S\'ersic profile to model and subtract from the \textit{HST} image the galaxies located within $\simeq 17$ arcsec from the BCG centre. The best--fitting parameters for the galaxy profiles are listed in Table~\ref{tab:galpar}; the corresponding galaxies  are indicated in  Fig.~\ref{fig:sysgen} with the same numbering sequence  as in Table~\ref{tab:galpar}.  Snapshots of the galaxy-subtracted \textit{HST} image are shown in Fig.~\ref{fig:snap}. 
\begin{table*}[!htb]
\caption[]{Best-fit parameters obtained by modelling the galaxy luminosity 
with the S\'ersic profile \citep{sersic1968}.}
\begin{center}
\begin{tabular}{llllllll}
\multicolumn{8}{c}{\textsc{Best-fitting S\'ersic profile parameters}} \\
 \noalign{\smallskip}\hline \noalign{\smallskip}
 ID &  $\rm x_{c}$ & $\rm y_{c}$ & Axis &  $r_{e}$  & Position  & S\'ersic  & $\rm  m_{606W}$ \\
 &deg(J2000) & deg(J2000) & ratio & [kpc] & Angle & index &   \\ 
\noalign{\smallskip}\hline
\hline\noalign{\smallskip}
BCG & $120.23678$ & $36.056572$ & 	 $0.70\pm0.01$ & 	 $38.54\pm0.27$ & 	 $132.5\pm0.1$ & 	 $3.0\pm0.1$ & 	 $17.0\pm0.1$ \\ 
$1$ & $120.23598$ & $36.054391$ & 	 $0.83\pm0.01$ & 	 $1.48\pm0.01$ & 	 $112.8\pm0.1$ & 	 $2.3\pm0.1$ & 	 $20.7\pm0.1$ \\ 
$2$ & $120.2357$ & $36.053776$ & 	 $0.92\pm0.01$ & 	 $1.02\pm0.01$ & $13.1\pm0.1$ & 	 $2.9\pm0.1$ & 	 $21.6\pm0.1$ \\ 
$3$& $120.23855$ & $36.061407$ & 	 $0.50\pm0.01$ & 	 $3.85\pm0.04$ & 	 $78.7\pm0.2$ & 	 $2.6\pm0.1$ & 	 $20.7\pm0.1$ \\ 
$4$ &$120.24051$ & $36.059402$ & 	 $0.67\pm0.01$ & 	 $3.28\pm0.07$ & 	 $80.6\pm0.5$ & 	 $4.7\pm0.1$ & 	 $20.8\pm0.1$ \\ 
$5$& $120.24255$ & $36.056744$ & 	 $0.84\pm0.01$ & 	 $1.35\pm0.02$ & 	 $128.4\pm1.4$ & 	 $4.6\pm0.1$ & 	 $21.4\pm0.1$ \\ 
$6$& $120.23639$ & $36.055754$ & 	 $0.90\pm0.01$ & 	 $0.64\pm0.01$ & 	 $131.6\pm4.0$ & 	 $2.3\pm0.1$ & 	 $22.0\pm0.1$ \\ 
$7$& $120.24148$ & $36.060146$ & 	 $0.79\pm0.01$ & 	 $2.32\pm0.04$ & 	 $61.7\pm0.9$ & 	 $5.1\pm0.1$ & 	 $21.1\pm0.1$ \\ 
\noalign{\smallskip}\hline
\end{tabular}

\end{center}
\label{tab:galpar}	     
\tablefoot{The algorithm \textit{GALFIT} is used.
The fit was performed using  the \textit{ACS} F606W filter image. 
We list here the galaxy ID (the galaxy numbering is the same used 
in Fig.~\ref{fig:sysgen}), the centroid coordinates, the axis ratio, 
the effective radius $r_e$, the  positional angle, the S\'ersic index 
$n$ and the galaxy integrated magnitude, computed  in the ST magnitude system. 
Here we report the $68\%$ statistical errors as determined from \textit{GALFIT}. 
The errors were rounded to the closest value greater than zero.
}
\end{table*}
\subsubsection{Mass profiles of cluster galaxies}
 In order to reduce the number of parameters to an adequate number we need to relate the  total mass of cluster galaxies  to an observable quantity; we will thus assume a scaling relation that links the mass of a galaxy to its luminosity. We will also assume that the morphological parameters of the underlying galaxy halo match the values inferred from the galaxy luminosity; this is a reasonable assumption because of the tight correlation between light and mass profile in  \mbox{early-type}  galaxies (see e.g. \citealp{gavazzi2007,koopmans2006}).
 For this reason,  the galaxy centroid, ellipticity and position angle were fixed to the  values  obtained from the analysis performed with \textit{SExtractor}  of the $r$ filter LBC images. 
 However, the   parameters derived from the  LBC  data for  some of the galaxies close to the lensed images or to the BCG appear to be incorrect, because of an inefficient de-blending for the lower spatial resolution of  LBC images. In these cases we adopted the  parameters (namely  centroid coordinates, position angle, major/minor axis length and magnitude) estimated through  \textit{SExtractor} from the \textit{HST} F606W image.  \\
  We parameterised the mass of cluster galaxies  with the dual   pseudo isothermal elliptical mass distribution (hereafter dPIE, \citealp{kassiola1993,kneib1996}), following  \cite{jullo2007,brainerd1996} and several other authors. The dPIE profile is given by
\begin{equation}
\label{eq:piemd}
\rho(r)=\frac{\rho_0}{(1+r^2/r_{\rm core}^2)(1+r^2/r_{\rm cut}^2)}.
\end{equation}
The dPIE cut radius can be roughly considered as  the half mass radius\footnote{The cut radius corresponds to the 3-D half mass radius for a null core radius \citep{eliasdottir2007}.}. 
The two-dimensional surface mass density distribution is obtained by integrating Eq.~\ref{eq:piemd} \citep[see also ][]{limousin2005}:\\
\begin{equation}
\label{eq:piemd_2d}
\Sigma(R)=\frac{\sigma_0^2 r_{\rm cut} }{2\mathrm{G}(r_{\rm cut}-r_{\rm core})}\left(\frac{1}{\sqrt{r_{\rm core}^2+R^2}}-\frac{1}{\sqrt{r_{\rm cut}^2+R^2}}\right),
\end{equation}
where $\sigma_{0}$ is a characteristic central velocity dispersion, which  cannot be simply related to the measured velocity dispersion \citep{eliasdottir2007}. \\
The free parameters in the dPIE model  are thus the core radius $r_{\rm core}$, the truncation radius $r_{\rm cut}$, and the characteristic velocity dispersion $\sigma_0$. For elliptical galaxies, the Faber-Jackson relation (hereafter FJ, \citealp{faber1976}) predicts that the central velocity dispersion $\sigma_{0}$  is proportional to  the  galaxy  luminosity $L$ following the scaling relation
\begin{equation}
\label{eq:scal1}
\sigma_0  =   \sigma_0^\star \left(\frac{L}{L^\star} \right)^{1/4},
\end{equation}
  \noindent where $L^\star$ is a given arbitrary galaxy luminosity  and  $\sigma_0^\star$ is  its corresponding dPIE velocity dispersion; our reference magnitude is $m^\star_{r}=18.0$.
We will assume the following scaling relations for the characteristic radii of the dPIE profile:
\begin{equation}
\label{eq:scal2}
\left\{ \begin{array}{l}
r_{\rm core}  =  r_{\rm core}^\star (L/L^\star)^{1/2}\;, \\
r_{\rm cut}  =   r_{\rm cut}^\star (L/L^\star)^\alpha\;;  \\
\end{array}
\right.
\end{equation}
here again $r_{\rm core}^\star$  and $r_{\rm cut}^\star$ are the dPIE scale parameters for a $L^\star$ galaxy.  The exponent of the scaling relation for  $r_{\rm cut}$ determines how the mass scales with respect to the light distribution. If $\alpha = 0.5$ the mass-to-light ratio ($ML$)  of the galaxy  is constant and independent
  of its luminosity, as usually assumed in strong and weak lensing analyses \citep{richard2007,hoekstra2004}. Assuming $\alpha = 0.8$ one obtains that the  mass-to-light ratio scales with luminosity according to $ML \propto L^{0.3}$ \citep{halkola2006,natarajan1997}. In this paper we will assume  $\alpha = 0.5$, but in $\S \ $~\ref{sec:slres} we will compare the results obtained under this assumption with those obtained by assuming $\alpha = 0.8$.  We did not test alternative scaling relations for the cluster galaxy core radius, since  we did not identify any reliable constraint on the galaxy inner  mass distribution. \\
 The galaxy parameters optimised in the lens modelling  are the central velocity dispersion and  the truncation radius, whereas the other dPIE parameters (ellipticity, position and orientation of the halo) were tied to the values inferred from the analysis of the galaxy light profile. \\
The normalisation  for the core radius scaling was set to $r_{\rm core}^\star$=0.035  arcsec $\simeq 0.15$ kpc, while for the velocity dispersion  normalisation $\sigma_0^\star$ we adopted the uniform prior [120.0--200.0] $\rm km\ \rm s^{-1}$. Following the relation between luminosity and velocity dispersion found by \cite{bernardi2003},  our reference magnitude corresponds to a velocity dispersion $\simeq$~170~$\rm km\ \rm s^{-1}$ for an  \mbox{early-type}  galaxy. The prior  $\sigma_0^\star\in [120.0-200.0]$~$\rm km\ \rm s^{-1}$ adopts a conservative upper limit; the lower limit was decreased to  $120.0$~$\rm km\ \rm s^{-1}$ to take into account the recent findings of \cite{natarajan2009} about the difference in the dPIE velocity dispersion between cluster galaxies in the core or in the outer regions. \\
The  truncation radius normalisation determines the steepness of the galaxy mass profile. A common and reasonable approach is to fix it to a likely value  (see R09 and \citealp{limousin2009}), since usually the strong lensing system does not provide enough constraints on the mass distribution of the small scale perturbers.  However, the mass distribution of  \mbox{early-type}  galaxies in high--density environment is  likely to be truncated compared to field galaxies with an equivalent luminosity \citep{halkola2007,limousin2007b,natarajan2002}. For this reason, the truncation of cluster galaxy halos could be different from one case to another.  \\
 Thanks to the peculiar  lensing system of Abell~611, an estimate of the  truncation radius normalisation will be derived in one dedicated tests (see  Sect.~\ref{sec:slres}).  
The perturbing galaxy labelled 1 in Fig.~\ref{fig:sysgen}  was optimised singularly, since the strong constraints on its Einstein radius provided by the conjugate systems B.4/B.5 enable a reliable estimate of its mass distribution. Because however neither R09 or \cite{newman2009}  unlinked this galaxy from the cluster member catalogue, we ran a test  to verify if and how this single perturber can affect the global test results. 
\subsection{Modelling of the BCG}
\label{sec:bcg}
\begin{figure}[!htb]
\begin{center}
\makebox[0.5\textwidth][c]{\includegraphics[width=0.4\textwidth]{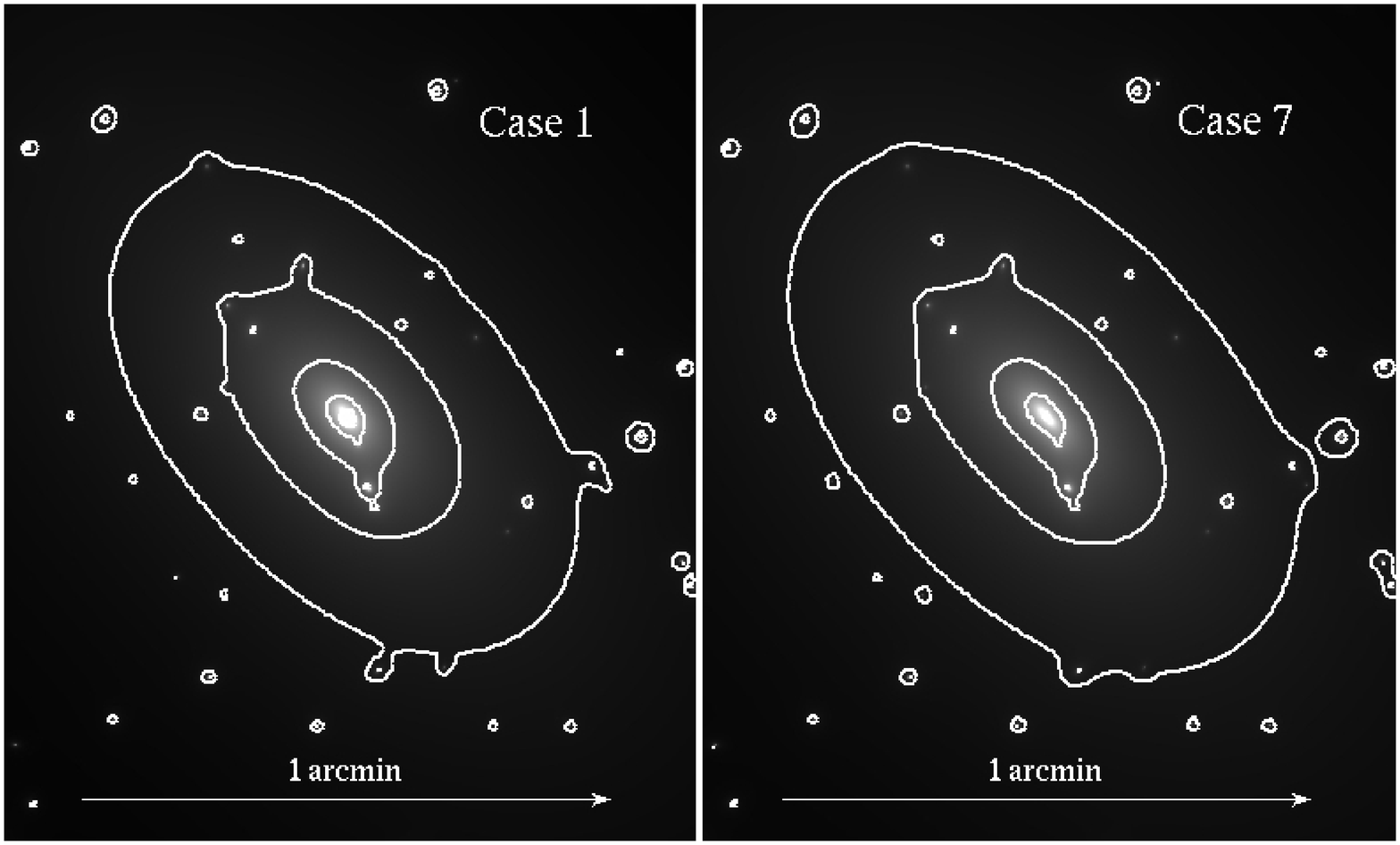}}\par
\makebox[0.5\textwidth][c]{\includegraphics[angle=-90,width=0.23\textwidth]{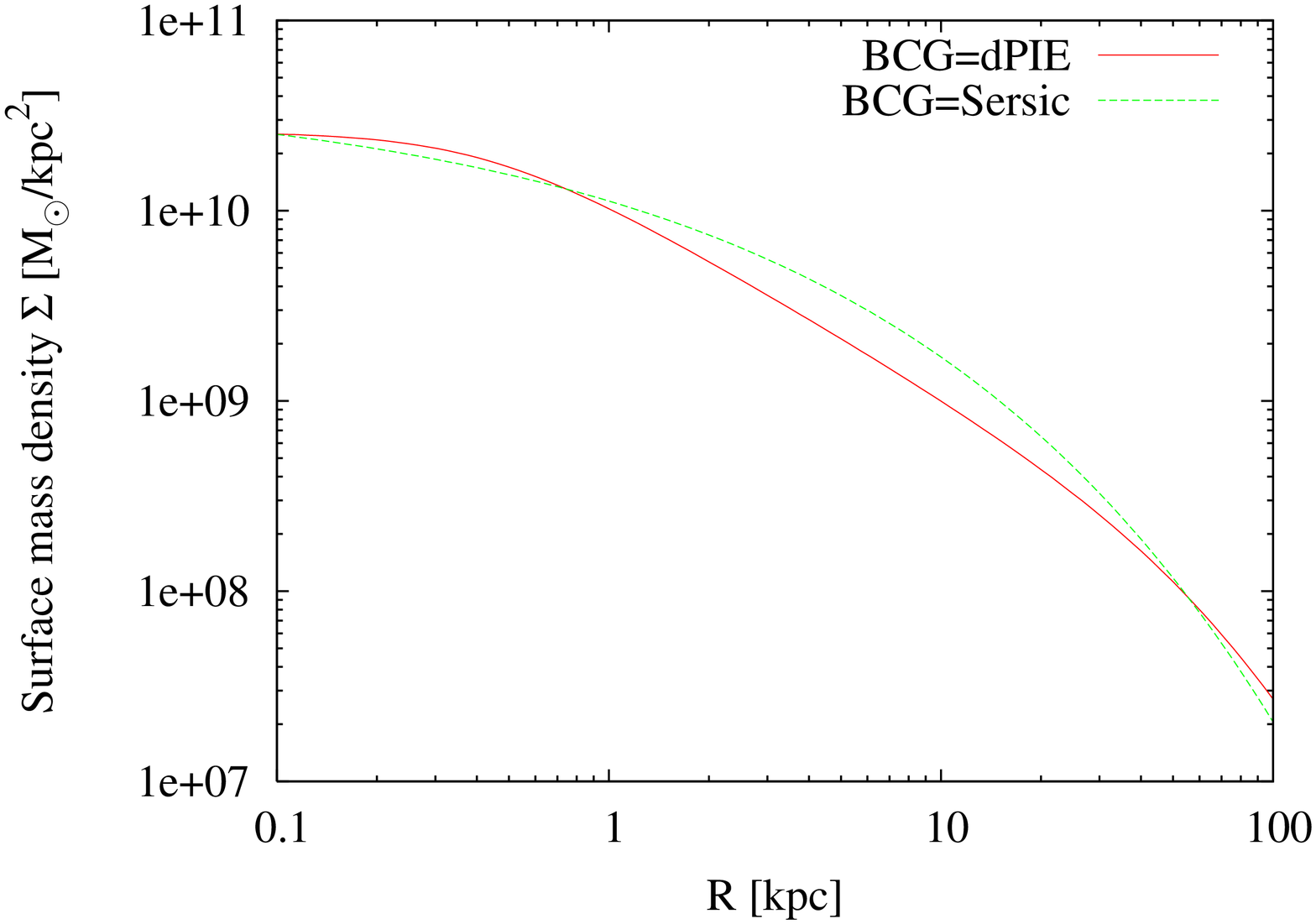}\includegraphics[angle=-90,width=0.23\textwidth]{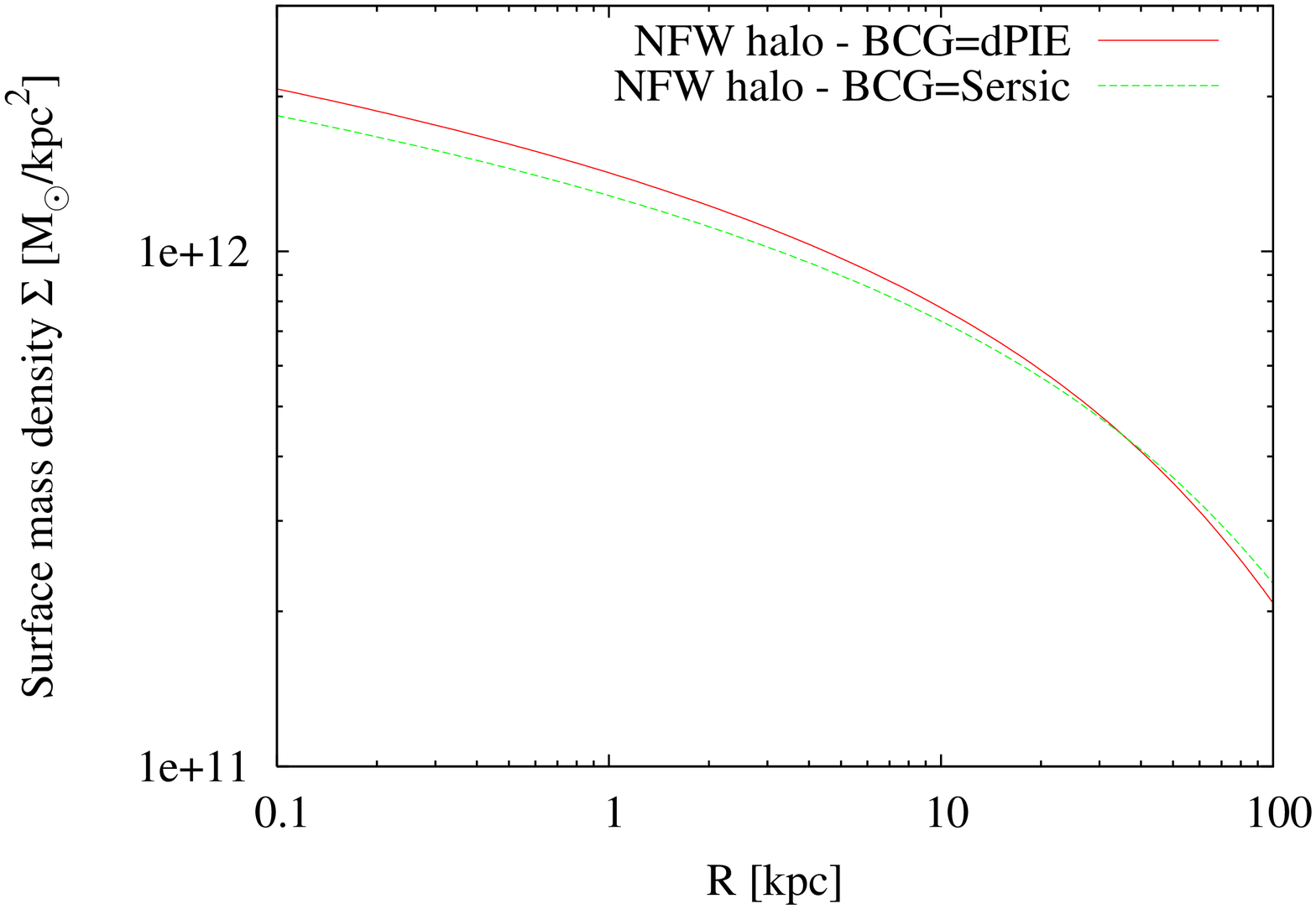}}
\end{center}
\caption[]{\textit{[Top Panels]} Comparison  between the  projected mass maps obtained when modelling the BCG with a  dPIE profile (left panel, case 1 in  Table~\ref{tab:tab_l1}) or  with a  S\'ersic profile  (right panel, case 7 in Table~\ref{tab:tab_l1}). See Table~\ref{tab:tab_l1} and the text for further details. \textit{[Bottom Panels]} Comparison between the surface mass density profiles for the BCG halo \textit{[left panel]} and the NFW cluster scale halo \textit{[right panel]} derived when modelling the BCG with a dPIE profile (red solid line, case 1) or with a S\'ersic profile (green dashed line, case 7).}
  \label{fig:bcgmass}
  \end{figure}

The BCG itself was included in the lens modelling as a dPIE halo.  However, although the BCG was parameterised with the same mass functional as the cluster members, its mass model refers only to its \textit{baryonic}  content, while for cluster galaxies what is actually modelled is the \textit{total} mass. As pointed out in the  work of \cite{miralda1995}  (see also \citealp{limousin2008,dubinski1998,mellier1993}), the dark matter halo of the BCG is indistinguishable  from  the cluster halo in the lens modelling.\\
 For this reason a   direct comparison of the measured central velocity dispersion of the BCG to the fitted  $\sigma_{0,\rm dPIE} $ would lead to incorrect conclusions: the former is determined by the combined contribution of the dark and baryonic matter, while the latter is a fitting parameter referred to as a  ``stellar-only'' BCG component. \\
 The dPIE mass parameters - characteristic central velocity dispersion  $\sigma_{0,\rm dPIE}$ and truncation radius $r_{\rm cut}$- were optimised for the BCG in the ranges     [100.0-350.0]~$\rm km\ \rm s^{-1}$ and  [8.0-31.0] arcsec ($\simeq [34.6-134.2]$ kpc), respectively: these ranges take into account the results obtained through a strong lensing analysis by R09 and by N09,  to allow a comparison with their results (see $\S \ $~\ref{sec:prev}).\label{page:bcgopt}  \\
     The  degeneracy between the central baryonic mass budget and the dark matter content with regard to the strong lensing results is currently well established. It is still to be discussed whether the choice of the BCG mass profile can impact the derived cluster mass parameters. To assess this point, we ran a dedicated test adopting  the S\'ersic profile to parametrise  the   BCG mass. 
      The S\'ersic mass profile  has the same functional form as Eq.~\ref{eq:sersic}, but the parameter $\Sigma_e$ refers to the mass density at the effective radius $r_e$.  \\
      For the modelling of the BCG  we adopted the morphological parameters derived from the two-dimensional fitting of the BCG luminosity profile obtained with \textit{Galfit}, listed in Table~\ref{tab:galpar}. We optimised  the   S\'ersic  index within $\pm 10\%$ of the best-fitting value, to take into account the measurement errors.  After some preliminary tests, we chose as lower limit to the    $\Sigma_e$ optimisation range the value of   $2 \times 10^8  M_{\odot}\rm kpc^{-2}$  to derive a total mass for the BCG similar to the  mass derived when modelling the BCG as a dPIE profile, for a safer comparison between the results  obtained in both cases (see the total projected mass maps and the surface mass density profiles shown in Fig.\ref{fig:bcgmass}).
\subsection{Strong lensing results}
\label{sec:slres}
We present  in this section the results of the  parametric analysis of the strong lensing system in Abell~611.
 In order  to assess the effect of the galaxy parameterisation on the best-fitting results for the cluster-scale halo we performed several tests, changing the manner of including the cluster galaxies. \\
  We list below the characteristics of the lens modelling for the cases  presented in Tab.\ref{tab:tab_l1}.
\begin{table*}[!htb]
\caption[]{Median values for the total mass parameters, obtained with the strong lensing analysis.}
\setlength{\extrarowheight}{5pt}
\begin{center} 
\resizebox{\textwidth}{!}{%
 \begin{tabular}{l@{\@{\hspace{0.5pt}}}lcccccccc@{\@{\hspace{1pt}}}c@{\@{\hspace{1pt}}}c}
 \multicolumn{12}{c}{\textsc{\large Results of the strong lensing analysis of Abell 611}} \\
 \hline  
 Case  &    &   $\rm x_{ c} $  &   $\rm y_{c}$  &   $ e$  &   $PA$   &   $\rm r_{\star}$ &   $\rm c_{200}$  &   $\rm M_{200}$  &   $\sigma_{0}$  &   $\chi^{2}_{\rm red}[d.o.f.] $  &   $\rm rms_i$\\
   &     &   [arcsec]  &    [arcsec]  &    &   [deg]  &   [kpc]  &    &   $[\rm 10^{14} M_{\odot}]$  &   [km $\rm s^{-1}$]  &     &  [arcsec] \\
 \hline 
 \multirow{4}{*}{1} 
  &   Cluster &   -0.6$\pm0.2$  &   0.9$\pm0.2$   &   0.25$\pm0.01$   &    132.8$\pm0.2$   &   214.3$\pm23.6$   &   6.78$^{+0.81}_{-0.37}$  &   4.68$\pm0.31$  &   1300.5$\pm15.7$   &    0.6 [61]  &   0.33\\
  &   BCG  &   (0.0)   &  (0.0)  &       &     &    85.5$^{+4.0}_{-67.8}$  &     &      &   310.9$^{+22.2}_{-40.0}$  &    &  \\
   &   Perturber 1 &   (2.33)   &  (-7.85)  &       &     &   22.1$^{+6.1}_{-10.5}$  &      &      &   176.3$^{+9.4}_{-6.8}$  &  \\
  &   Ref. Galaxy  &       &     &       &     &   37.1$^{+6.4}_{-14.0}$  &      &      &    179.1$^{+15.9}_{-11.9}$ &   \\[2pt]
  \hline
  \multirow{4}{*}{2} 
  &   Cluster &   -0.6$\pm0.1$  &   0.9$\pm0.2$  &   0.25$\pm0.01$  &    132.8$\pm0.2$  &   228.0$^{+16.6}_{-20.8}$   &   6.43$^{+0.56}_{-0.32}$   &   4.87$\pm0.27$  &   1307.2$\pm16.4$   &    0.6 [62]  &   0.37\\
  &   BCG  &  (0.0)  &   (0.0) &       &     &    91.1$^{+8.4}_{-55.0}$  &     &      &   324.7$^{+4.3}_{-33.1}$   &   &   \\
   &   Perturber 1  &    (2.33)  &  (-7.85) &       &     &   23.1$^{+4.4}_{-10.7}$   &      &      &   178.2$\pm6.5$ &   \\
   &   Ref. Galaxy  &       &     &       &     &   (43.3)  &      &      &    187.1$^{+1.5}_{-16.7}$ &   \\[2pt]
  \hline
 \multirow{4}{*}{3} 
  &   Cluster &   -0.6$\pm0.1$  &   0.9$\pm0.2$  &   0.25$\pm0.01$  &    132.9$\pm0.2$  &   218.3$^{+22.7}_{-16.9}$   &   6.63$^{+0.67}_{-0.32}$   &   4.68$^{+0.33}_{-0.18}$  &   1297.4$\pm15.0$   &    0.6 [62]  &   0.32\\
  &   BCG  &    (0.0)  &  (0.0) &       &     &    96.2$^{+13.2}_{-43.4}$  &     &      &   311.5$^{+14.3}_{-36.8}$   &   &   \\
   &   Perturber 1 &    (2.33)  &  (-7.85) &       &     &   23.8$^{+1.1}_{-14.3}$   &      &      &   176.0$\pm6.5$  &  \\
   &   Ref. Galaxy  &       &     &       &     &   (43.3)  &      &      &    175.9$\pm11.8$  &  \\
 \hline
 \multirow{4}{*}{4}  
 &   Cluster &   -0.3$\pm0.1$  &   0.4$\pm0.1$  &   0.21$\pm0.01$  &    132.9$\pm0.2$  &   158.6$^{+8.8}_{-5.3}$   &   8.75$^{+0.19}_{-0.42}$   &   4.01$\pm0.13$  &   1303.7$\pm7.0$   &    0.9 [63]  &   0.50\\
  &   BCG  &   (0.0)   &  (0.0) &       &     &    96.0$^{+2.1}_{-52.3}$  &     &      &   132.8$^{+47.1}_{-2.2}$   &   &   \\
   &   Perturber 1 &     (2.33)  &  (-7.85)  &       &     &     &      &      &     &  \\
   &   Ref. Galaxy  &       &     &       &     &   48.7$^{+0.6}_{-4.8}$ &      &      &    198.6$^{+0.2}_{-2.3}$ &   \\[2pt]
  \hline
 \multirow{7}{*}{5}
  &   Cluster &   -0.4$\pm0.1$  &   0.9$\pm0.2$  &   0.23$\pm0.01$  &    132.8$\pm0.2$  &   192.0$\pm18.7$   &   7.40$\pm0.60$   &   4.40$\pm0.22$  &   1295.4$\pm11.5$   &   1.0 [74]  &   0.31\\
  &   BCG  &   (0.0)  &  (0.0)  &       &     &    99.4$^{+7.0}_{-48.4}$  &     &      &   263.0$\pm35.6$   &    &  \\
   &   Perturber 1 &     (2.33)  &  (-7.85)  &       &     &  12.0$^{+7.6}_{-1.9}$   &      &      &   180.4$^{+1.3}_{-14.0}$ &     &  \\
   &   Ref. Galaxy  &       &     &       &     &   (43.3) &      &      &    190.5$^{+5.6}_{-9.7}$  &   &  \\
    	  \cline{5-8}
   	 \multicolumn{11}{c}{Redshift estimate of the candidate system D} \\
    	\cline{5-8}
   	 &    &     &   &  $ z_{D.1}$   &  $ z_{D.2}$   &  $ z_{D.3}$    &  $ z_{D.4}$   &    &    &   \\
    	 &    &    &   &  2.07$\pm0.04 $ &    2.32$\pm0.26$ &   2.43$\pm0.39$  &  2.09$\pm0.05$  &    &    &  \\
  \hline 
   \multirow{9}{*}{ 6}
  &   Cluster &   -0.5$\pm0.2$  &   0.8$\pm0.2$  &   0.25$\pm0.01$  &    133.2$\pm0.4$  &   211.8$^{+30.2}_{-11.1}$    &  6.81$^{+0.69}_{-0.41}$   &   4.59$^{+0.34}_{-0.16}$  &   1294.1$\pm13.9$   &   0.7 [52]  &   0.20\\
  &   BCG  &   (0.0)   &   (0.0) &       &     &    95.6$^{+24.1}_{-28.9}$  &     &      &   297.2$^{+30.4}_{-46.8}$   &    &  \\
   &   Perturber 1  &    (2.33)   &  (-7.85)  &       &     &  19.4$^{+16.8}_{-4.5}$   &      &      &   186.4$^{+15.7}_{-10.2}$ &     &  \\
    &   Perturber 2  &    (3.14) &  (-10.05) &       &     &  15.9$^{+5.5}_{-8.3}$   &      &      &   108.2$^{+33.6}_{-22.4}$ &     &  \\
    &   Perturber 3  &    (-5.15) &  (17.42) &       &     &  11.4$^{+8.8}_{-2.1}$   &      &      &  125.4$^{+25.5}_{-16.2}$ &     &  \\
    &   Perturber 4  &    (-10.88) &  (10.22) &       &     &  16.2$^{+17.7}_{-0.8}$   &      &      &  103.5$^{+14.5}_{-7.7}$ &     &  \\
       &   Perturber 5  &    (-16.79) &  (0.60) &       &     &  14.6$^{+12.5}_{-1.8}$   &      &      &  80.9$^{+21.7}_{-10.8}$ &     &  \\
       &   Perturber 6  &    (1.13) &  (-2.78) &       &     &  14.8$^{+10.4}_{-3.9}$   &      &      &  153.8$^{+10.1}_{-64.0}$ &     &  \\
   &   Ref. Galaxy  &       &     &       &     &   (43.3)  &      &      &    149.6$\pm17.9$  &   &  \\
   \hline 
   \multirow{4}{*}{7}
  &   Cluster &   -0.8$\pm0.1$  &   1.0$\pm0.2$  &   0.20$\pm0.01$  &    132.8$\pm0.3$  &   286.9$^{+24.9}_{-11.4}$    &  5.59$^{+0.18}_{-0.31}$   &   6.32$^{+0.51}_{-0.23}$  &   1392.1$\pm19.0$   &   1.1 [61]  &   0.44\\[2pt]
  \cline{2-12}
  &   BCG$_{\rm Sersic}$  &   (0.0)   &   (0.0) &       &     &  $\Sigma_{e}$=2.08$^{+0.12}_{-0.01}$ $[10^8 \rm M_{\odot}/\rm kpc^2]$  &     &      &   S\'ersic index $n$=2.87$^{+0.26}_{-0.01}$   &    &  \\[2pt]
      \cline{2-12} 
   &   Perturber 1   &      (2.33)   &  (-7.85)  &       &     &  21.8$^{+1.6}_{-16.5}$   &      &      &   176.0$^{8.1}_{-6.0}$ &     &  \\
   &   Ref. Galaxy  &       &     &       &     &   36.5$^{+9.1}_{-10.8}$   &      &      &    161.7$^{+11.3}_{-20.9}$  &   &  \\[2pt]
    \hline
 \end{tabular}

}
\end{center}
\label{tab:tab_l1}  
\tablefoot{
The results are derived assuming an elliptical NFW profile for the cluster-scale lens component and an elliptical dPIE for the galaxy scale halos. We report only the values that were optimised in the strong lensing fit; additionally, the coordinates of the cluster galaxies are listed in round brackets for the sake of clarity, but they were not optimised. The columns list the coordinates of the halo centroid (expressed in arcseconds with respect to the BCG centre: at the cluster redshift and for the assumed cosmology, 1 arcsec=4.329 kpc), the potential ellipticity (here expressed as $e=(a^2-b^2)/(a^2+b^2)$), the halo position angle, the scale radius r$_\star$ (that indicates the scale radius  $r_{\rm s}$ for the cluster-scale NFW halo and the truncation radius $r_{\rm cut}$ for the galaxy-scale dPIE halos), the concentration parameter $c_{200}$,   the total mass $ M_{200}$,  the  characteristic velocity dispersion $\sigma_{0}$ for the NFW or the dPIE profiles, expressed in $\rm km\ s^{-1}$, the reduced $\chi^{2}$ (the number of degrees of freedom is enclosed within  square brackets) and the  root-mean-square  computed in the image plane  ($\rm rms_i$). The uncertainties represent the $68\%$ statistical errors, derived assuming a Gaussian probability distribution. If the upper and lower errors are significantly asymmetric we report both of them.  For a description of the cases listed here, see Pag.~\pageref{des}. }
\end{table*}  
\begin{figure*}[!ht]
 \begin{center}
\includegraphics[angle=-90,width=0.9\textwidth]{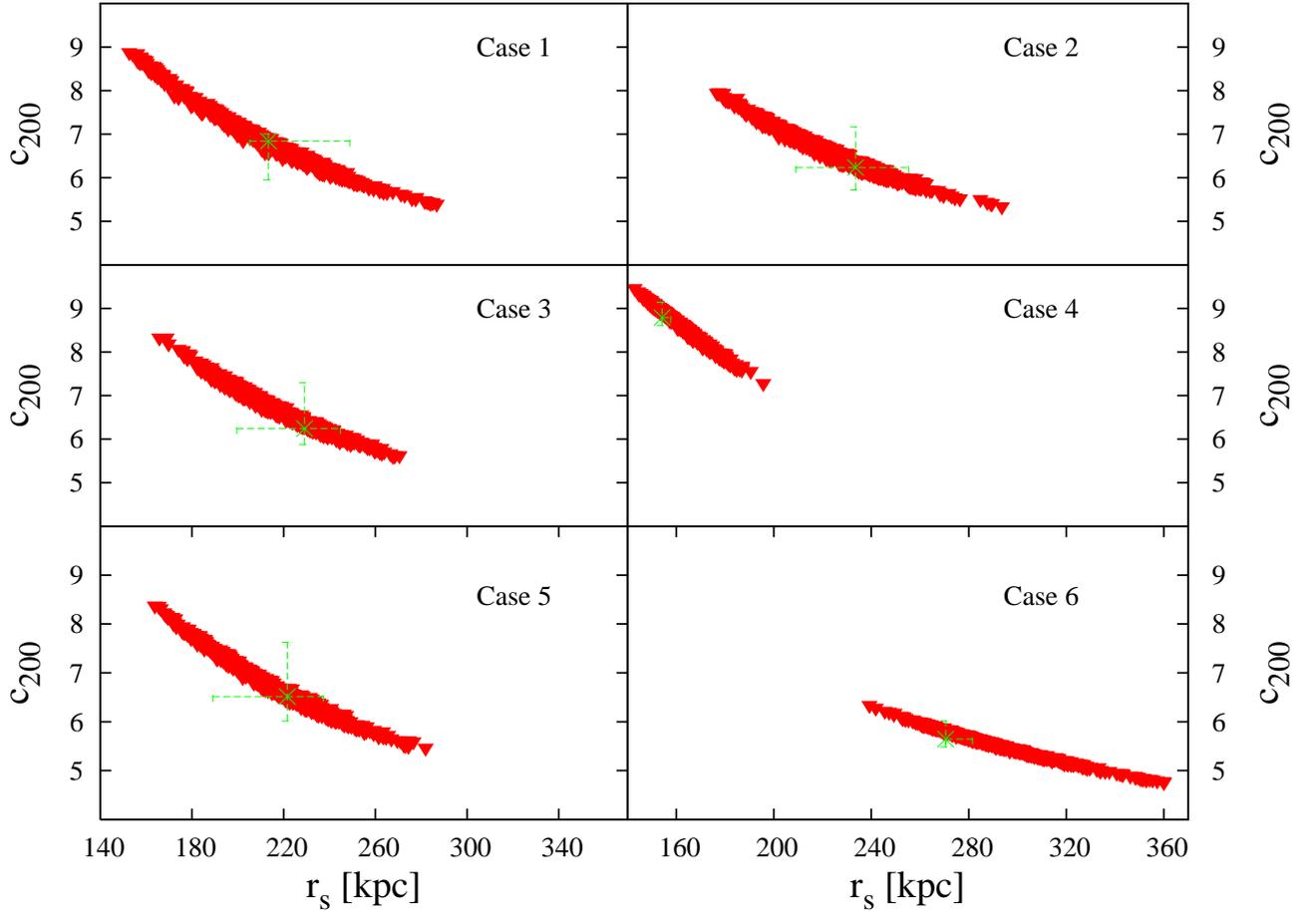} 
\end{center}
\caption{Scatter plots for the NFW parameters $r_{\rm s}$ and $c_{200}$ obtained for some representative cases.  The numbering of the cases corresponds to the numbering in Table~\ref{tab:tab_l1}. The red  point distribution marks the $3\sigma$ confidence level  for the two parameters. The green stars indicate the best-fit values; the error-bars refer to the $1\sigma$ statistical uncertainty.  We used the same knot systems as inputs and the same priors on the cluster halo mass distribution in all the six cases; the only differences were the modelling and the parameterisation of the cluster galaxies and of the BCG. 
The probability distribution in case 4 appears squeezed because it  crushed on the lower limit for the BCG characteristic central velocity dispersion. 
The lens model results for the cases shown here are listed in Table~\ref{tab:tab_l1}. }
 \label{fig:histog1}
\end{figure*}
  \begin{itemize}\label{des}  
  \item   \emph{Case 1:}    both the reference truncation radius and the  reference central velocity dispersion for the cluster--galaxy scaling relation were optimised  in the ranges [4 -- 12] arcsec and [120.0 -- 200.0] $\rm km\ \rm s^{-1}$, respectively.\\ 
  In this case,  and when not otherwise stated, the mass parameters of galaxy n.~1 in Fig.\ref{fig:sysgen} were individually optimised (the optimization ranges were \mbox{$r_{\rm cut}\in$ [0.5 -- 8.0] arcsec} and \mbox{$\sigma_0\in$ [90.0 -- 190.0] $\rm km\ \rm s^{-1}$)} and the exponent  $\alpha$  in Eq.~\ref{eq:scal2} was set equal to 0.5. 
 \item  \emph{Case 2:}  the reference central velocity dispersion was optimised, whereas  the reference  truncation radius was fixed to the value of $r_{\rm cut}^\star$=10 arcsec. The exponent  $\alpha$  in Eq.~\ref{eq:scal2} was assumed to be equal to 0.8 (therefore  the Mass to Light ratio $ML$ is $\propto 0.3$).
  \item  \emph{Case 3:} as in case 2, but we assumed that $\alpha=0.5$  (therefore $ML$ is constant and independent of the galaxy luminosity).
 \item  \emph{Case 4:}  a constant $ML$ scaling relation was assumed for all cluster galaxies. This is the only case in which    the galaxy  n.~1 in Fig.\ref{fig:sysgen} was not individually optimised.
 \item  \emph{Case 5:} we adopted the same lens model as in case 3, but we included as constraint the candidate lens system individuated by R09 and by N09, labelled D in  Fig.\ref{fig:sysgen} and in Tab.\ref{tab:input} in the Appendix.
  \item  \emph{Case 6:} a constant $ML$ scaling relation was adopted for  all the cluster galaxies, with the exception of those that can be tested through the strong lensing analysis, for their direct effect on some lensed images (galaxies 1 to 6 in Fig.\ref{fig:sysgen}), which were individually optimised.
 \item  \emph{Case 7:}  the same lens model as in case 1 was assumed, but the BCG was modelled with an elliptical S\'ersic profile.
 \end{itemize}
   It is not redundant to verify whether the mass distributions of the  perturbing galaxies  differ from  the distribution that can be derived adopting a scaling relation, since, as shown in Table~\ref{tab:galpar},  these galaxies have significantly different light profiles. Moreover, the intrinsic scatter in the FJ scaling relation between velocity dispersion and luminosity for  \mbox{early-type}  galaxies is not taken into account  in the strong lensing modelling.
    We stress that in all the aforementioned cases, except for case 7, the BCG was modelled as a dPIE halo (see Sect.~\ref{page:bcgopt} for more details on the BCG modelling). \\
The results of the  strong lensing analysis of Abell~611 for the aforementioned cases are listed in Table~\ref{tab:tab_l1}.  The critical and the caustic lines  and the images predicted in all tested  cases  are shown in the  Fig.\ref{fig:resl} in the Appendix.
Evidently depending on the different lens modelling the results on the cluster total mass can change by up to $15\%$, so the  parameterisation of the perturber galaxies and of the BCG has a \mbox{non-negligible} impact on the strong lensing results; the degeneracy between  cluster and galaxy mass parameters is shown in Fig.\ref{fig:histog1}.
 In particular, a strong  degeneracy is found between the fiducial BCG central velocity dispersion and the cluster concentration/total mass  (see the top panels in Fig.\ref {fig:degen}). \\
It is remarkable that the resulting NFW cluster parameters depend  not only on the BCG baryonic mass, but also on the BCG mass profile.  In case 1 and in case 7 we show the results obtained by modelling the BCG with a dPIE or with a  S\'ersic profile, respectively; the corresponding projected mass maps and surface mass density profiles are shown in Fig.~\ref{fig:bcgmass}.  The  BCG mass derived in both cases is similar, but the inferred cluster parameters are significantly different. In particular, the cluster total mass is larger and the NFW concentration parameter lower when modelling the BCG with a S\'ersic profile. This result can be  explained by considering that, in this case, the surface mass density of the  best-fitting S\'ersic profile in the BCG central region is larger compared with the  best-fitting dPIE profile (see the bottom left panel in Fig.~\ref{fig:bcgmass}). The inferred concentration for the cluster scale halo in the latter case will thus be higher because of the degeneracy between the cluster halo and the BCG  mass with respect to the lens total mass, which is the quantity constrained by the gravitational lensing.\\
  However,  the  estimates of the \emph{total}  mass in the central regions (where the strong lensing constraints are found) derived in the different cases mutually agree, (see the discussion in Sect.~\ref{sec:comparison}); the modelling of the galactic component appears to have a negligible impact on the total mass results in the central region. Conversely, the differences among several of the test cases arise when the cluster mass is extrapolated outside the region that is accessible through the strong lensing analysis.   \\ This   is another evidence that the modelling of the BCG stellar mass  can directly affect the cluster total mass results: in order to obtain  reliable mass estimates through  a parametric strong lensing analysis, it is crucial to break the degeneracy between the different lens components.\\
Despite the difference in the best-fitting results, all models return  probability regions in the  parameter space that are marginally consistent with each other. As previously found for the \mbox{X-ray} analysis, the strong lensing results appears relatively stable against possible systematic errors, which in this case are mainly constituted  by modelling uncertainties. \\
As stated before, the perturber galaxy n.1 (see Fig.~\ref{fig:sysgen}) was individually optimised in all the tests that we ran, except in case 4, due to the tight constraint on its mass distribution provided by the conjugated  images B.4 and B.5. We found  a mild degeneracy between the mass of this galaxy, the mass of the stellar content of the BCG, and the concentration of the NFW halo, as expected given the  proximity of this perturber to the BCG (see the bottom panels in Fig.~\ref{fig:degen}).  \\
\begin{figure}[!htb]
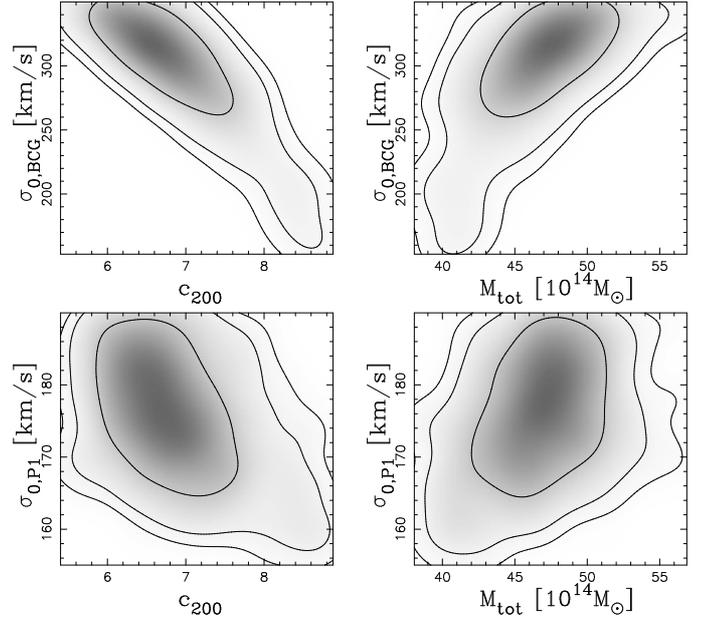

\begin{center}
\begin{tabular}{@{\@{\hspace{1pt}}}lr@{\@{\hspace{1pt}}}}
\includegraphics[width=0.23\textwidth]{14120fg09a.ps} & 
\includegraphics[width=0.23\textwidth]{14120fg09b.ps} \\
 \includegraphics[width=0.23\textwidth]{14120fg09c.ps} & 
\includegraphics[width=0.23\textwidth]{14120fg09d.ps} \\ 
\end{tabular}
\end{center}
\caption[]{Scatter plots of  cluster--halo parameters ($c_{200}$ and $M_{\rm tot}$) versus  parameters of the modelling of either the BCG  or the perturber galaxy n.1 (hereafter P1, see text for details). Here we show the posterior probability distributions obtained in a representative case (case 1 in Tab.\ref{tab:tab_l1}). The plotted parameters are: \begin{inparaenum} \item the  NFW concentration parameter $c_{200}$  versus the BCG  fiducial central velocity dispersion $\sigma_{0,\rm BCG}$ \textit{[top left panel]};  \item the  cluster halo  mass $M_{\rm tot}$ versus  $\sigma_{0,\rm BCG}$ [\textit{top right panel}]; \item the  NFW concentration parameter $c_{200}$ versus the  fiducial central velocity dispersion $\sigma_{0,\rm P1}$ of the galaxy P1 [\textit{bottom left panel}]; \item  the  cluster halo  mass $M_{\rm tot}$ versus $\sigma_{0,\rm P1}$  [\textit{bottom right panel}].  \end{inparaenum}  The contours show the 68\%, 95\%  and 99\% confidence levels.  The colour code refers to the value of the $\chi^2$ estimator.}
\label{fig:degen}
\end{figure}
Moreover, in case 6 we optimised singularly the galaxies whose mass distribution can be constrained through the strong lensing effect alone. As it is evident from Fig.~\ref{fig:cut_sig}, optimising  in wide ranges  the dPIE parameters for these perturber galaxies  returns, as expected, large uncertainties in the mass parameters.
\begin{figure}[!htb]
\begin{center}
 \includegraphics[angle=-90,width=0.5\textwidth]{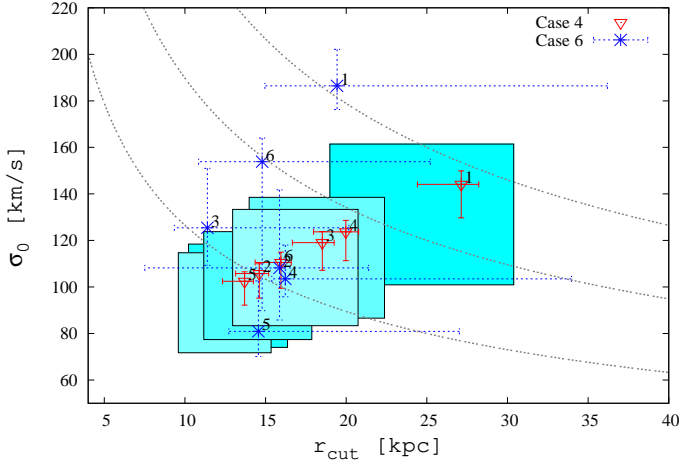}
\end{center}
\caption[]{Median values of the dPIE truncation radius $ r_{\rm cut}$ vs the characteristic dPIE central velocity dispersion $\sigma_{0}$ for the galaxies that perturb the strong lensing system of Abell~611. The red triangles show the results obtained linking the galaxy parameters to the same scaling relation  in the lens modelling (Case 4 in  Table~\ref{tab:tab_l1}), while the blue stars represent the results obtained optimising the galaxies individually  (Case 6 in  Table~\ref{tab:tab_l1}). The grey dashed lines show the  iso-density contours. The error bars indicate the 1$\sigma$ uncertainty, while the cyan boxes indicate the  3$\sigma$ confidence region for the results which refer to Case 4. }
 \label{fig:cut_sig}
\end{figure}
However, for some perturbers  the parameters inferred through an individual optimisation appear to be incongruous with those predicted  assuming the  parameter scaling derived from the Faber-Jackson relation.  It is reasonable to question whether neglecting the scatter associated to the galaxy parameter scaling relations in the strong lensing  analysis could  bias the inferred results.\\
 It seems likely that in  most cases  this bias, if any, can be safely negligible, as it is obvious by comparing case 1 to case 6. Nonetheless, in the presence of galaxies that strongly affect the strong lensing features, the biasing  effect on the final results could be more significant (again, this is the case  of the perturber galaxy labelled n.1  in this work, as can be seen by comparing case 1 to case 4). \\
 Regarding the strong lensing systems, all our models predict that the tangential arc (system A) originates by the large deformation of a single source located on a ``naked cusp'', i.e. a tangential caustic extending outside the radial caustic \citep{kormann1994,bartelmann1998,rusin2001}.
 Naked cusps are a quite rare caustic configuration which mainly occurrs in highly elliptical lens systems whose surface density is only moderately above the critical surface density required for the production of multiple images (``marginal lenses'', \citealp{blanford1987,bartelmann2000}), with a few naked cusps  detected so far \citep{lewis2002,oguri2008}. Sources placed within the exposed tangential caustic form only three magnified images on the same side of the lens: this explains why the tangential arc in Abell~611 has no  counter-image. \\
The position and the shape of the observed images  are well reproduced by all our models, as is obvious from the low scatter both in the image and in the source plane resulting from the best-fit models:  all tests returned  low values of the reduced $\chi^2$ ($\chi^2_{\rm red} \leq 1$) and of the image plane root-mean-square  ($\rm rms_i \leq 0.5$ arcsec) (see also the fitting results for a representative case - Case 1 - listed in Tab.\ref{tab:chi2l}). \\
In particular, the lens models of cases 4 and 7 (see Tab.\ref{tab:tab_l1}) predict  a central  image for the  system B in a position almost coincident with a radial-oriented feature lying close to the centre of the BCG (see Fig.\ref{fig:rad}  and the predicted central image in Fig.\ref{fig:resl}; this central image was also indentified and interpreted by N09).   However, since the detection of this feature   is not certain, we did not include it as an additional constraint.\\
Regarding the third system C, our identification of the  positive parity counter-image, which follows R09, appears reliable owing to the very low scatter resulting for this system. Our models predict three additional counter-images: two of them should be located close to the BCG and to the galaxy n.1, and are predicted to be demagnified with respect to the identified image, so they are likely not detectable. A negative parity counter-image is predicted at about 2-3 arcsec (depending on the different lens models) above the observed one.
\begin{figure}[!htb]
\begin{center}
\includegraphics[width=0.4\textwidth]{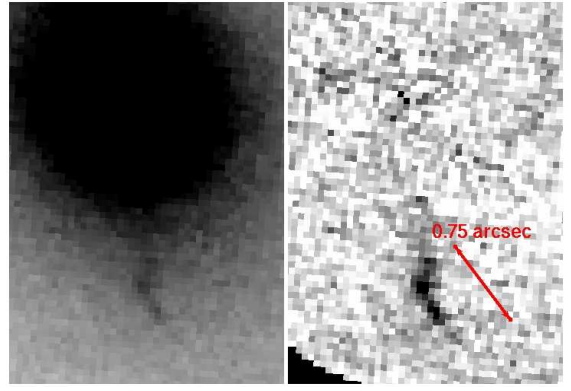} 
\end{center}
\caption{\textit{ACS/HST} snapshot of a radial-oriented feature, likely a central counter-image corresponding to  System B; the BCG is visible [subtracted] in the left [right] panel.}
 \label{fig:rad}
\end{figure}
\begin{figure}[!htb]
\begin{center}
 \includegraphics[height=0.2\textheight]{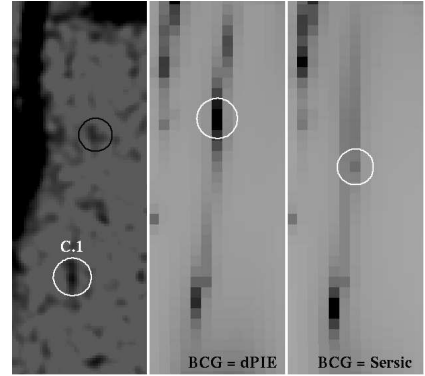}
 \end{center}
 \caption{Comparison between the \textit{ACS/HST} F606W observation \textit{[left panel]}, smoothed with a 3 FWHM pixel Gaussian  filter for a better visualisation of  the faint lensed image C.1 (marked with a white circle), and the reconstructed image systems predicted from the lens models in case 1  (\textit{[middle panel]}: in this case the BCG was modelled as a dPIE potential) and case 7 (\textit{[right panel]}: the BCG was parametrised with a S\'ersic profile - see Table~\ref{tab:tab_l1}). The white circles in the middle and right panels mark the centre of the predicted counter-images. The black circle in the left panel indicates a candidate counter-image of C.1, indentified with SExtractor. The three images are WCS-aligned; the flux scale of the second and the third panel is the same. }
 \label{fig:c_comp} 
\end{figure}
 In that area there are no evident counter-images, but a faint one, identified with SExtractor (S/N $\simeq$ 3.9), whose position could be marginally compatible with the location of the predicted image. The flux ratio derived when modelling the BCG as a dPIE halo does not match the observed one well, while the images  predicted when modelling the BCG with a S\'ersic   profile follow  the observed flux ratio more closely (see Fig.~\ref{fig:c_comp}), because of the difference in the  inferred mass for the galactic perturbers. 
\begin{table}[htb]
\caption[]{Fitting results of the source--plane optimisation.}
\begin{center}
\begin{tabular}{l|ccc}
 \multicolumn{4}{c}{\textsc{Goodness of fit}} \\
\hline 
System & $\chi^2$ & $\rm rms_s ['']$ & $\rm rms_i ['']$ \\
\hline \hline   
A.1 &1.73 &0.067 &0.30\\
A.2 &5.56 &0.082 &0.92\\
A.3 &   2.98& 0.085& 0.40\\
A.4& 0.23 &0.024 &0.09\\
A.5 &0.15 &0.014 &0.28\\
A.6 &1.27 &0.050 &0.12\\
A.7 &2.88 &0.066 &0.38\\
A.8 &0.98& 0.051  &0.16\\
\hline
B.1& 2.80 &0.109& 0.29\\
B.2& 2.44 &0.100 &0.29\\
B.3& 2.37 &0.092 &0.19\\
B.4& 2.23 &0.088 &0.22\\
B.5& 2.56& 0.083 &0.19\\
\hline
C.1 &0.58 &0.052 &0.10\\
\hline
Total & $\chi^2$ [d.o.f] & $\rm rms_s ['']$ & $\rm rms_i ['']$ \\
      &  28.76 [61] &  0.079&  0.33 \\  
\hline
\end{tabular}

\end{center}
\label{tab:chi2l}  
\tablefoot{We show here the goodness-of-fit indicators  for a representative case (case 1 in Table~\ref{tab:tab_l1}).  The columns list the knots system ID, the total $\chi^2$ for  the  system, the root-mean-square  computed in the source plane ($\rm rms_s$), and  in the image plane ($\rm rms_i$). The last row reports the total $\chi^2$ estimator, referring to a model with 61 degrees of freedom. }
\end{table}  
 This evidence would seem to favour the modelling of the BCG with    a S\'ersic   profile  over the   dPIE profile: but we aim to highlight that other explanations could be possible to the observed mismatch in the flux ratios for the images of the third system. For example, the effect of unresolved substructures along the line of sight  would cause   the dimming of the negative-parity image in cusp systems \citep{keeton2003} (see also \citealp{oguri2008}). Another explanation could be the misidentification of the counter-image C.2, which would cause its predicted magnification to be incorrect.\\
The  candidate system D  looks plausible considering the low scatter in the observed and predicted positions. The redshift estimates of the four knot systems are marginally consistent. However,  the results of the cluster  mass parameters are not significantly changed by  including  this system. 
\section{Comparison between X-ray and strong lensing results}
\label{sec:comparison}
\begin{figure*}[htb]
\begin{center}
\begin{tabular}{ll}
\includegraphics[angle=-90,width=0.48\textwidth]{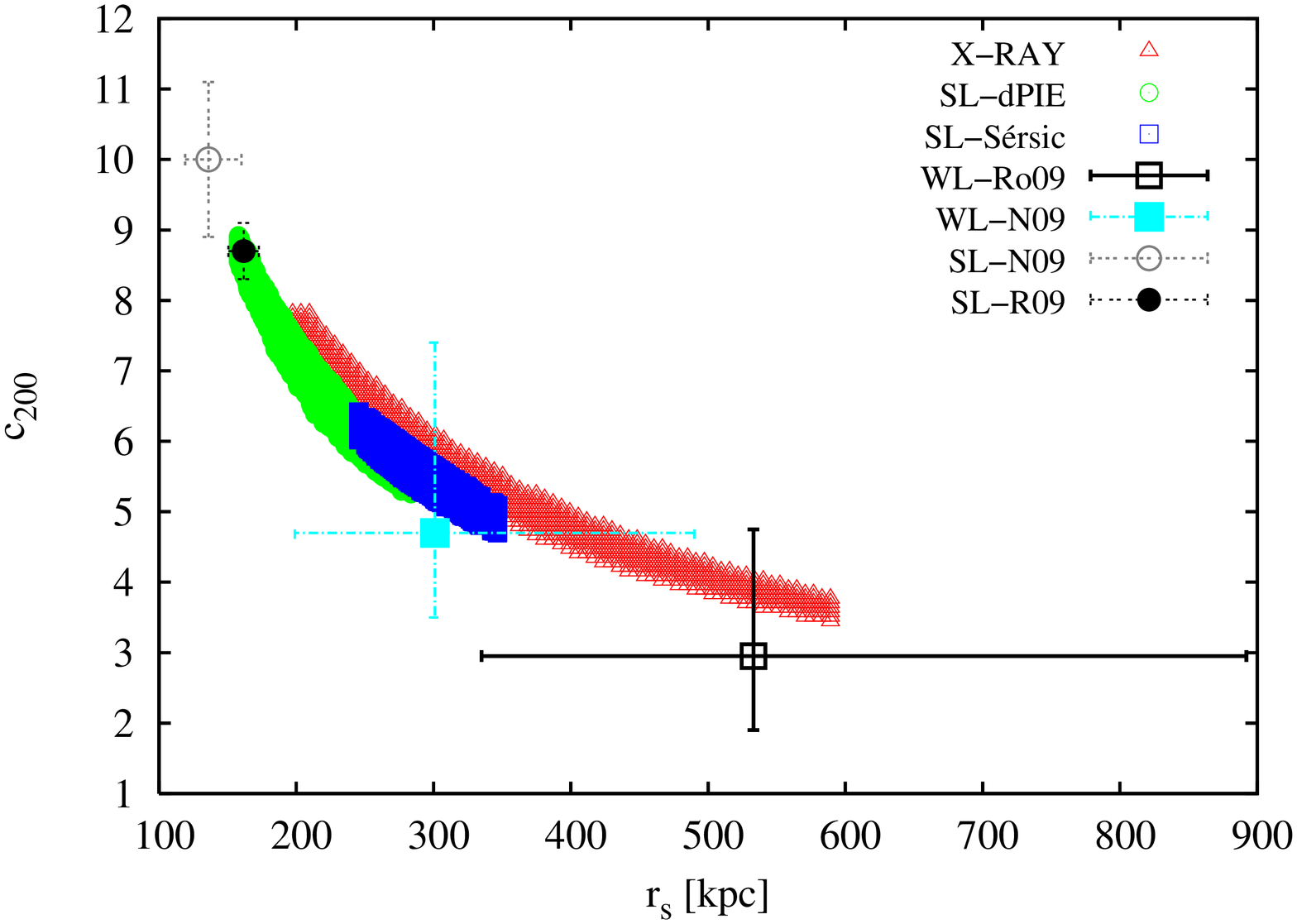} & \includegraphics[angle=-90,width=0.48\textwidth]{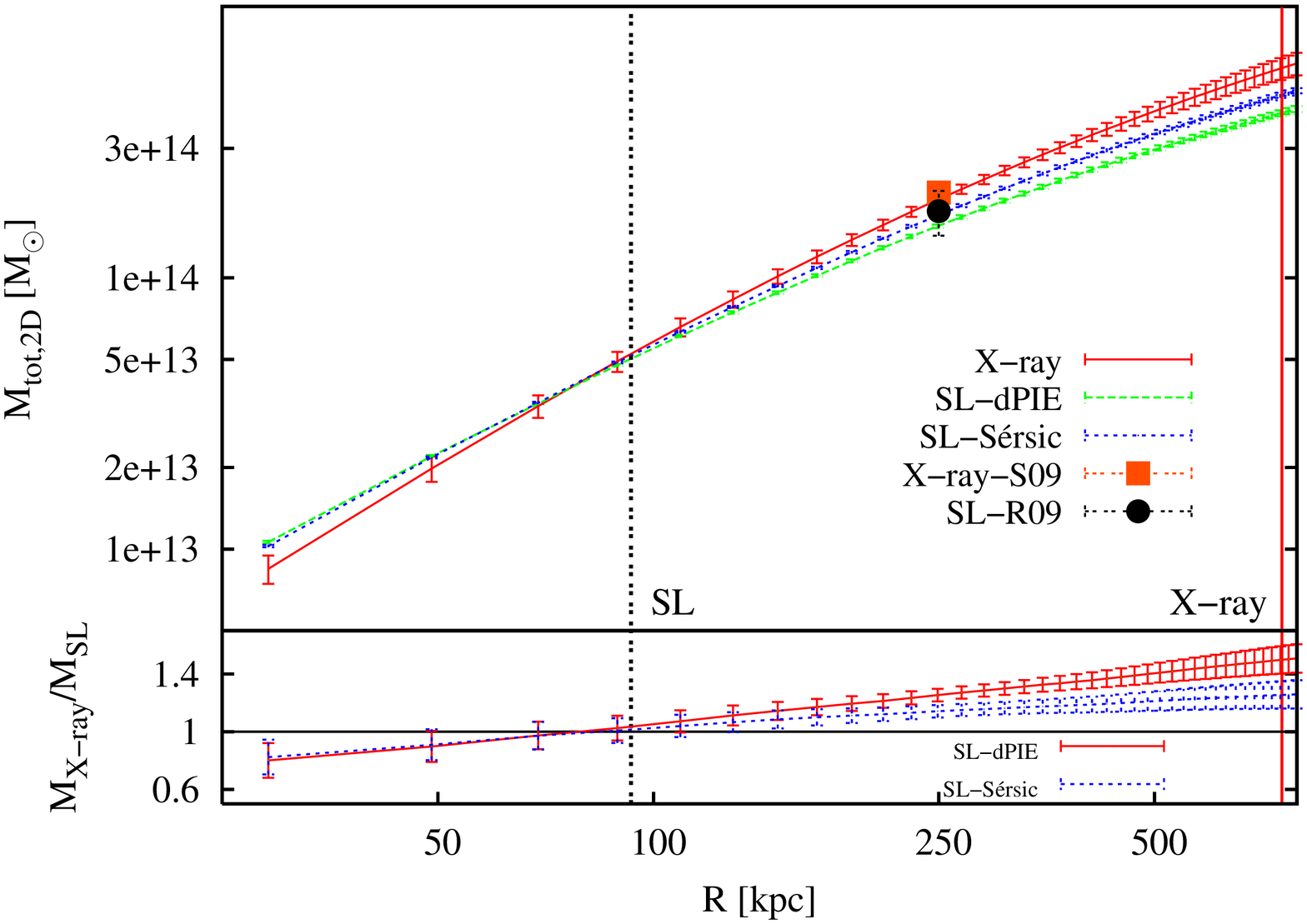} 
\end{tabular}
\end{center}
\caption{\textit{[Left panel]} Comparison of the NFW parameter probability distribution obtained with our  \mbox{X-ray} and strong lensing analyses. The \mbox{X-ray}  and  the strong lensing   contours show the 3-$\sigma$ confidence level (the strong lensing results were marginalised over the other fit parameters). The \mbox{X-ray} contours were obtained deriving the \mbox{X-ray} background from the blank-sky data-set, while the strong lensing result refers to case 1 and to case 7 in Table~\ref{tab:tab_l1}.  For an easier comparison we overplotted some previous recent results. We show in the plot the results obtained through a strong lensing analysis by \cite{richard2009} (label SL-R09),  through a weak and a strong lensing analysis by \cite{newman2009} (labels WL-N09 and SL-N09, respectively) and in Paper I through a weak lensing analysis (label WL-Ro09; the results shown here were obtained by adopting a spherical NFW profile; the shear catalogue in this case was derived through the shapelet decomposition technique). The  bars show the  1-$\sigma$ errors. \textit{[Top right panel]} Projected 2-D mass profiles obtained with the \mbox{X-ray}  (red solid line) and with the strong lensing analyses when modelling the BCG as a dPIE  (green dashed line) or  a S\'ersic halo (blue dashed line). The vertical lines mark the spatial range of the strong lensing (black dotted line) and \mbox{X-ray} (red solid line) constraints. The \mbox{X-ray} [strong lensing] profile was derived by measuring the enclosed mass in cylinders centred on the \mbox{X-ray} emission centroid [BCG]. The distance between the  \mbox{X-ray}  centroid and the BCG centre is $\simeq 1\pm1.5$ arcsec. The  mass map derived from the strong lensing analysis  was extrapolated up to a radius of $200$ arcsec from the mass measurements obtained in the central regions. The statistical errors on the strong lensing profiles were determined computing the standard deviation from 300 mass map realizations. As a comparison, we report the estimates of the projected total mass within 250 kpc obtained by \citealp{richard2009} through a strong lensing analysis (label SL-R09, marked with a filled black circle) and through an \mbox{X-ray} analysis by \citealp{sanderson2009a}  (label X-ray-S09, marked with a filled orange square).  \textit{[Bottom right panel]} Ratios between the \mbox{X-ray} and the strong lensing projected mass estimates;  the strong lensing estimates were obtained   by modelling the BCG through a dPIE  (red solid line) or  a S\'ersic  (blue dotted line) profile. }
 \label{fig:res_comp}
\end{figure*}

We will compare here the results obtained with the strong lensing and the \mbox{X-ray} analyses for the galaxy cluster Abell~611. The left panel in Fig.~\ref{fig:res_comp}  shows the probability distributions of the NFW profile parameters obtained with both analyses, while the right panel compares the corresponding projected mass profiles. For the strong lensing analysis we show  the $3\sigma$ region, marginalised over the other fit parameters, derived in case 1 and in case 7 of Table~\ref{tab:tab_l1}.   The best-fit values for the NFW profile parameters and for the corresponding  mass  are listed in Table~\ref{tab:tab_both}. \\
To unfold this issue, higher resolution observations of the  Sunyaev-Zel'dovich effect measurements towards this cluster would play a crucial role, because they would allow us with the \mbox{X-ray} data to estimate the projected axis ratio along the line of sight. 
Obviously the  best-fit values for the  scale radius and the  concentration parameter  derived with the \mbox{X-ray} and the strong lensing analyses are  consistent within a  $2-\sigma$ range; however, the results derived from the strong lensing analysis in case 7  better agree with the \mbox{X-ray} estimates. In particular, the strong lensing mass estimates for Abell~611 are lower than the  mass derived through our \mbox{X-ray} analysis which is lower by up to a factor of two.  \\
\begin{table}[!htb]
\caption[]{Comparison of the mass results obtained 
with the \mbox{X-ray} and the strong lensing analyses. }
\begin{center}
\begin{tabular}{lccc}
\\
\multicolumn{4}{c}{\textsc{Comparison  between X-ray and Strong Lensing Results}} \\
\hline
 & $\rm r_{s}$ & $\rm c_{200}$ & $\rm M_{200}$ \\
 & [kpc] & & [$10^{14}\rm M_{\odot}$] \\[3pt]  
\hline
\hline\\[-5pt]
X-ray - \emph{[Blank sky]}  & $350.3\pm 79.6$ & $5.18\pm 0.84$ & $9.3\pm 1.4$     \\[2pt]
Strong lensing -\emph{[case 1]} & $214.3^{+22.2}_{-24.3}$  & $6.78^{+0.81}_{-0.37}$  & $4.7\pm0.3$\\[2pt]
Strong lensing -\emph{[case 7]} & $286.9^{+24.9}_{-11.4}$  & $5.59^{+0.18}_{-0.31}$  & $6.3\pm0.4$\\[3pt]
\hline
\end{tabular}   
        
\end{center}
\label{tab:tab_both}  
\end{table} 
This result is somehow surprising, since so far the discrepancies found between strong lensing and \mbox{X-ray} mass estimates have the opposite trend, with the smaller \mbox{X-ray} masses (see e.g. \citealp{richard2009b,peng2009,gitti2007,med2007}).  However, the  mass estimates for Abell~611 derived in  three representative cases agree very well in the  region where the strong lensing constraints are found (approximately $R \leq 100$ kpc), as is obvious from the right panel of Fig.~\ref{fig:res_comp}. On the other hand, the strong lensing mass is smaller than the \mbox{X-ray} one when extrapolating the mass profile in the cluster peripheral regions.\\
 This result confirms that strong lensing provides a robust  estimate  of the \textit{total} mass of the cluster (DM plus baryonic component) in the  central region; it moreover suggests that the disagreement between the \mbox{X-ray} and the strong lensing masses can be related to the incorrect estimate of the total baryonic mass  budget (for example, the reader can compare the results obtained in case 4 and in case 1 in Table~\ref{tab:tab_l1}) or of its radial profile, because  the resulting cluster mass distribution will be incorrectly steepened (with a higher concentration parameter and a smaller scale radius)  to account for the missing baryonic mass component.\\
  While this possible source of systematic errors in the strong lensing mass estimate was already observed in other studies, the  dependence of the  derived cluster total mass on the assumed BCG mass profile is a new evidence in the strong lensing analysis of real galaxy clusters, which was already observed through the analysis of simulated galaxy clusters by \cite{meneghetti2010}. A similar but less significant effect is related to the scaling of the $ML$ relation, which is assumed to derive the cluster galaxy masses, as is evident by comparing case 2 and case 3  in Table~\ref{tab:tab_l1}. \\
Another explanation of the X-ray/strong lensing mass mismatch for Abell~611 could be an oblate-like  cluster geometry: as demonstrated in \cite{gavazzi2005} (see also  \citealp{morandi2010a,corless2009,morandi2010b,meneghetti2010}), the orientation of the halo along  the plane of the sky would cause the lensing mass to under-estimate the true one. On the contrary, the \mbox{X-ray} derived masses are the most stable against  possible projection effects because of the cluster asphericity (see again \citealp{gavazzi2005}). \\ 
\section{Comparison with previous studies}
\label{sec:prev}
 A combined analysis of \mbox{X-ray} data and of Sunyaev-Zel'dovich effect measurements was performed by  \cite{defilippis2005}  to recover the three-dimensional potential distribution  for a sample of 25 massive clusters. They estimated that the elongation of  Abell~611 along the line of sight  is  $e_{\rm los}\simeq \ 1.05 \pm 0.37$, while its maximum axial ratio is  $\simeq \ 1.19\pm 0.27$, so neither an oblate nor a prolate orientation of Abell~611 can be excluded. \\
\cite*{morandi2007a} and \cite{morandi2007b} analysed  \textit{Chandra} \mbox{X-ray} data  for a sample of 24 \mbox{X-ray} luminous clusters to derive the scaling relations between their \mbox{X-ray} properties and the Sunyaev-Zel'dovich  effect measurements (in \citealp{morandi2007a}) and the profiles and the scaling properties of their  gas entropy (in \citealp{morandi2007b}). They analysed the  same  \textit{Chandra} observation of Abell~611 as we (see Table~\ref{tab:obs} for more details on the \textit{Chandra} dataset) by adopting a technique similar to the method  followed here.  Their estimate of the  total mass of Abell~611 is $ M_{500} = (5.18 \pm2.15) \times 10^{14} M_{\odot}$ at the radius  $ R_{500} =  (1.1\pm 0.2)$ Mpc which agrees very well with our \mbox{X-ray} result of $ M_{500} = (6.77 \pm 0.84) \times 10^{14} M_{\odot}$  at the radius  $ R_{500} =  (1.2\pm 0.1)$ Mpc. \\
A very recent re-analysis of the \textit{Chandra} data for Abell~611 is presented in \cite{sanderson2009a} (see also  \citealp{sanderson2009b,sanderson2009c}). The mass estimate for Abell~611 at  $ R_{500} =  (1.3\pm 0.1)$ Mpc derived in their work is $ M_{500} = (9.23 \pm2.8) \times 10^{14} M_{\odot}$, which again agrees well with our results. \\
An analysis of  the weak gravitational lensing effect induced by this cluster was performed by \cite{dahle2006}  through observations  in the \textit{I} and \textit{V} bands  obtained with the University of Hawaii Telescope and the Nordic Optical Telescope. Assuming an NFW profile to model the radial density distribution and a concentration parameter as predicted by \cite{bullock2001}, they obtained a mass value of $M_{180}  = (5.21 \pm 3.47) \times 10^{14} M_{\odot}$ within $R_{180}$. The same data were reanalysed by \cite{pedersen2007},  who derived an estimate of $M_{500} = (3.83 \pm 2.89) \times 10^{14} M_{\odot}$. \\
A very recent  study of the weak lensing effect in Abell~611 is presented in  Paper I. This work takes advantage of the imaging data  obtained through the Large Binocular Camera (LBC) mounted at the Large Binocular Telescope (LBT), which were presented in Sect.~\ref{sec:sl}. Their weak lensing analysis benefits by the wide field of view of the LBC  ($\simeq 23\ \rm arcmin \times 25\ \rm arcmin$) and two different methods of extracting the shear signal, the well known KSB approach \citep*{kaiser1995} and the  shapelet decomposition technique,  proposed more recently by \cite{refregier2003a,refregier2003b} and \cite{massey2005}. 
 Their  model-indepent  estimates of the total projected mass  within a radius of $\simeq 1.5$ Mpc are $M_{\rm projec,WL,1} = (7.7 \pm 3.3) \times 10^{14} M_{\odot}$ and
$M_{\rm projec,WL,2} = (8.4 \pm 3.8) \times 10^{14} M_{\odot}$ when adopting the  KSB  or the  shapelet method, respectively. These results agree very well with our \mbox{X-ray} mass estimate, i.e. $ M_{\rm projec,X-ray} = [9.8 \pm 1.4] \times 10^{14} M_{\odot}$ at  the radius R $\simeq$ 1.5 Mpc, while our strong lensing mass estimate is lower (see the discussion in the previous section and the right panel of Fig.~\ref{fig:res_comp}). \\
They also derived a model-dependent estimate by assuming that the total mass profile of the cluster can be parameterised through the NFW profile. The best-fitting results were a total cluster mass of $ M_{200} = 5.6^{+4.7}_{-2.7} [5.9^{+2.2}_{-1.7}] \times 10^{14}\  M_{\odot}$, at the radius $ R_{200} = 1.54^{+0.34}_{-0.31} \ [1.57^{+0.18}_{-0.17}]$ Mpc,  and a concentration parameter of  $c_{200}$ = $3.13^{+4.67}_{-1.74} [2.95^{+1.80}_{-1.05}]$ when adopting the KSB [Shapelets] approach. \\
For an easier comparison, the best-fitting values of the cluster-scale NFW profile derived in the former case are overplotted in Fig.~\ref{fig:res_comp}. These results seems to agree better with our  \mbox{X-ray} mass results as regards the strong lensing estimates. \\
Recently, \cite{richard2009} (R09)  published the results of a strong lensing analysis performed on a subsample of 10 massive clusters extracted from the Local Cluster Substructure Survey (LoCuSS), which included Abell~611. \\
 Another study dedicated to the analysis of the matter distribution in Abell~611 was recently performed  by \cite{newman2009} (N09). They reconstructed the  cluster mass distribution  by combining weak and strong lensing analyses and dynamical data, taking advantage of Subaru, \textit{HST},  and Keck data, respectively. The strong lensing parametric analyses of both R09 and N09 were performed with the software \textit{Lenstool}. We overplotted their results for an NFW profile fit in  Fig.~\ref{fig:res_comp}.\\
\cite{richard2009} considered  as constraints to the lens optimisation both the confirmed  (labelled A, B, and C in our notation) and the candidate (labelled D) systems (as in the case 5 in Table~\ref{tab:tab_l1}), but including only one set of conjugated knots for the tangential arc (system A in our notation). Their results are consistent with our strong lensing findings, which are obtained modelling the BCG as a dPIE halo within the $3\sigma$ confidence range, while  a  larger mismatch is found with our \mbox{X-ray} results and with the strong lensing results in case 7. The discussion in $\S \ $~\ref{sec:comparison} could likely also be applied to a comparison between the strong lensing results of R09 and the \mbox{X-ray} and weak lensing mass estimates of Abell~611. \\
Regarding the results obtained by N09 through the fit of a NFW profile to the strong lensing data, their  best-fitting values show an even higher discrepancy with our mass results: for example, they derive a  concentration parameter  significantly higher than the concentration derived from both our \mbox{X-ray} and our strong lensing analyses. An extensive comparison of our results with the findings of R09 and N09 is beyond the aims of this comparative section. One of the differences in the strong lensing modelling that could have an impact on the final results is the number of adopted constraints, because N09 included only the brightest knots of the systems A, B, and D in their strong lensing analysis.
 It could be that  using as input all or most of the constraints that can be derived from the morphology of the lensed images in Abell~611  could decrease the discrepancy between their results and ours. Indeed, it is apparent that their results follow the  degeneracies  expected in the strong lensing parameter estimates between the cluster halo parameters and the central stellar component, which can be reduced by forcing the best-fitting lens model to  reproduce realistic reconstructions of the observed lensed images. Moreover, it is interesting that the SL  best-fit parameters reported  in N09 agree better with the results we obtained by linking all the perturber cluster galaxies to the same  relation (case 4 in Table~\ref{tab:tab_l1}). This evidence reinforces our conclusions about the significant impact that the modelling of the galactic component can have in the strong lensing analysis of Abell~611, and it could suggest that adopting the same scaling relation for the cluster galaxies in this case could be inadequate.\\
The weak lensing results of N09, instead, agree within the $3\sigma$ errors with both our strong lensing and \mbox{X-ray} mass estimates.
\FloatBarrier
\section{Conclusions and discussion}
\label{sec:concl}
We presented new mass estimates for the relaxed cluster Abell~611, derived  through the  analysis of the cluster \mbox{X-ray}  emission and of its exceptional strong lensing system. We performed a non-parametric reconstruction of the gas density and temperature profile through the only  \textit{Chandra} observation available to date.  The cluster total mass was then estimated adopting the NFW profile as mass functional, under the assumptions of spherical symmetry and of hydrostatic equilibrium. \\
Our \mbox{X-ray} estimates of the cluster total mass are $ M _{200} = 9.32\pm 1.39 \times 10^{14}\  M_{\odot}$ and  $ M _{200} = 11.11\pm2.06  \times 10^{14}\  M_{\odot}$ when deriving the \mbox{X-ray} background through the blank-field observations or through the source dataset, respectively, thus demonstrating the effect of the \mbox{X-ray} background treatment on the cluster mass estimate. The inclusion of the metallicity as free parameter in the spectral fitting was found to have a negligible impact on the cluster mass results. \\
We reconstructed the projected mass distribution in the central region of Abell~611 through a parametric analysis of its remarkable strong lensing system, using the publicly available software \textit{Lenstool}.  We used as constraints to the lens optimisation a detailed system of conjugated lensed images, indentifying 14 sets of knots in three different lensed systems. We adopted an elliptical NFW profile to parametrise the smooth cluster halo; moreover we added the small-scale perturbers constituted by the cluster galaxies by adopting a scaling relation that links their total mass to their luminosity.  We ran several tests to verify if and how the strong lensing cluster  mass estimate  is dependent on the modelling of the galactic  component, in particular on the  distribution of the  BCG baryonic mass component. \\
 We found that the extrapolated strong lensing mass for Abell~611 is \begin{itemize}  
\item moderately sensitive to the mass modelling of the galaxy-scale perturbers that  determine the peculiar characteristics of the  strong lensing system in Abell~611; indeed, we obtained that the different cluster galaxy parametrisations can induce a scatter of up to $~10\%$ on the total mass estimate (when extrapolating the SL mass up to $R_{200}$) with respect to the mean  value (comparing Cases 1 to 6 in Table~\ref{tab:tab_l1});
\item   more significantly  affected by the inferred BCG baryonic mass budget (because of the degeneracy between the cluster halo  mass and the BCG baryonic mass)  and by the assumed  mass profile for the BCG baryonic halo. The deviation of the extrapolated total mass with respect to the mean value can be as high as $~40\%$ when adopting different mass profiles for the BCG stellar mass distribution (comparing cases 1 to 7 in Table~\ref{tab:tab_l1}).
\end{itemize}
For example,  from the strong lensing analysis we derived for Abell~611 the  mass estimates of  $ M _{200,\rm dPIE} = 4.68\pm0.31  \times 10^{14}\  M_{\odot}$, when modelling the BCG with a dPIE model, and $ M _{200,\rm Sersic} = 6.32^{+0.51}_{-0.23}  \times 10^{14}\  M_{\odot}$, when assuming the S\'ersic profile as the BCG mass functional. The baryonic BCG mass inferred in both cases is similar, and  both estimates are dependent on the optimisation limits imposed to the BCG mass parameters. \\
Our \mbox{X-ray} and strong lensing mass estimates agree well in the central regions, where the total projected mass can actually be constrained through the strong lensing analysis, while  we find a significant difference between the \mbox{X-ray}  mass and the extrapolated SL mass in the cluster outer region. This result suggests that the observed  mismatch between the \mbox{X-ray} and the strong lensing mass  could be caused by the incorrect estimate of the BCG stellar mass content or of the total mass of cluster galaxies in the central regions, because gravitational lensing alone cannot separate the mass of the different lens components. \\
However, even with these caveats in the interpretation of the results, the \mbox{X-ray} and strong lensing mass estimates for Abell~611 in the most opportune cases are consistent within the $3\sigma$ error range. These findings support the hypothesis that taking into account the possible sources of  errors and  deepening  our knowledge of the current limitations of the \mbox{X-ray} and strong lensing analyses, both techniques return robust results. Their reliability can be  increased even more by extending the detailed analysis of  possible systematic biases affecting the mass estimates to other appropriate objects, i.e. massive, relaxed galaxy clusters like Abell~611.
\section*{Acknowledgements}
These results are based on observations made with the NASA/ESA Hubble Space telescope, with the Chandra \mbox{X-ray} Observatory and with the Large Binocular Telescope. We thank the anonymous referee for carefully reading the article, and for useful comments and suggestions that improved the presentation of our results. We thank Jean Paul Kneib and the \emph{Lenstool} developers for making their lensing software public.  AD  acknowledges the support of European Association for Research in Astronomy (MEST-CT-2004-504604 Marie Curie - EARA EST fellowship).  AD, SE and MM  acknowledge the financial contribution from contracts ASI-INAF I/023/05/0, I/088/06/0 and I/016/07/0. LF and MR acknowledge the support of the European Commission Programme 6-th framework, Marie Curie Training and Research Network ``DUEL'', contract number MRTN-CT-2006-036133. LF is partly supported by the Chinese National Science Foundation Nos. 10878003 \& 10778725,
973 Program No. 2007CB 815402, Shanghai Science Foundations and Leading Academic Discipline Project of Shanghai Normal University (DZL805).

\bibliography{master2}

\onecolumn

\appendix
\section{Multiple--image systems}
\begin{figure*}[!htb]
\begin{center}
 \hbox{\includegraphics[width=0.3\textwidth]{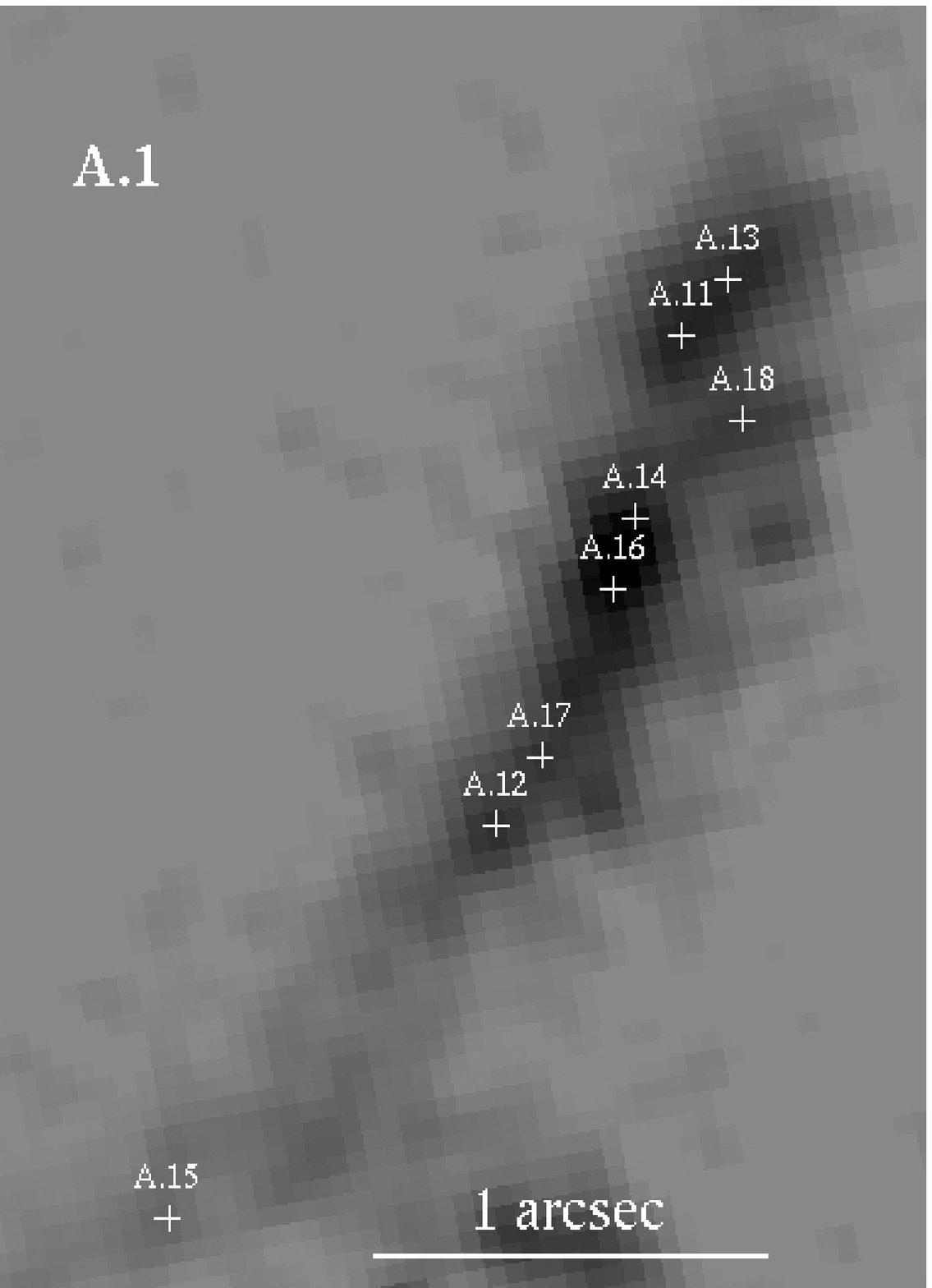}\hspace{0.4cm}\includegraphics[width=0.3\textwidth]{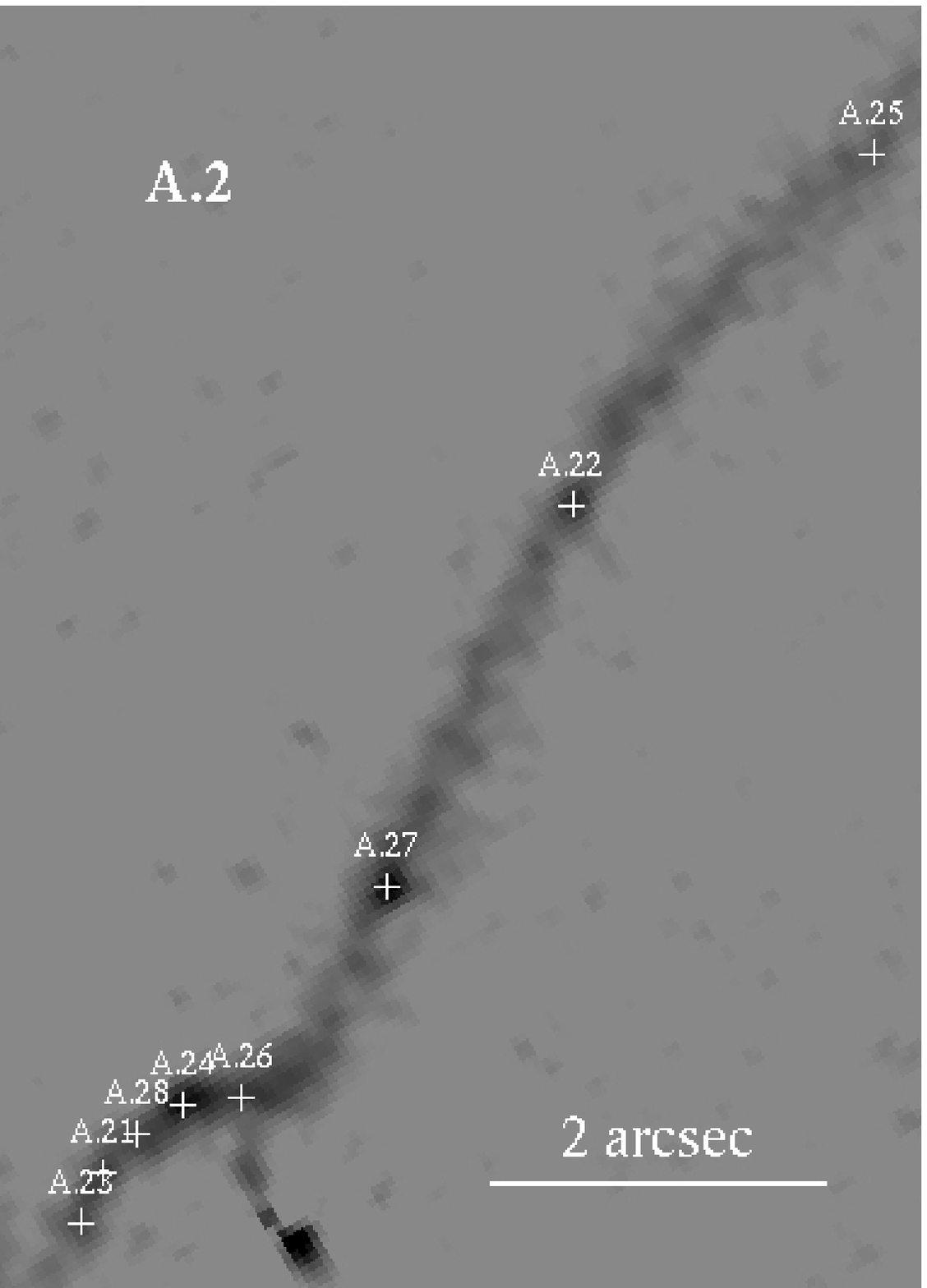}\hspace{0.4cm}\includegraphics[width=0.3\textwidth]{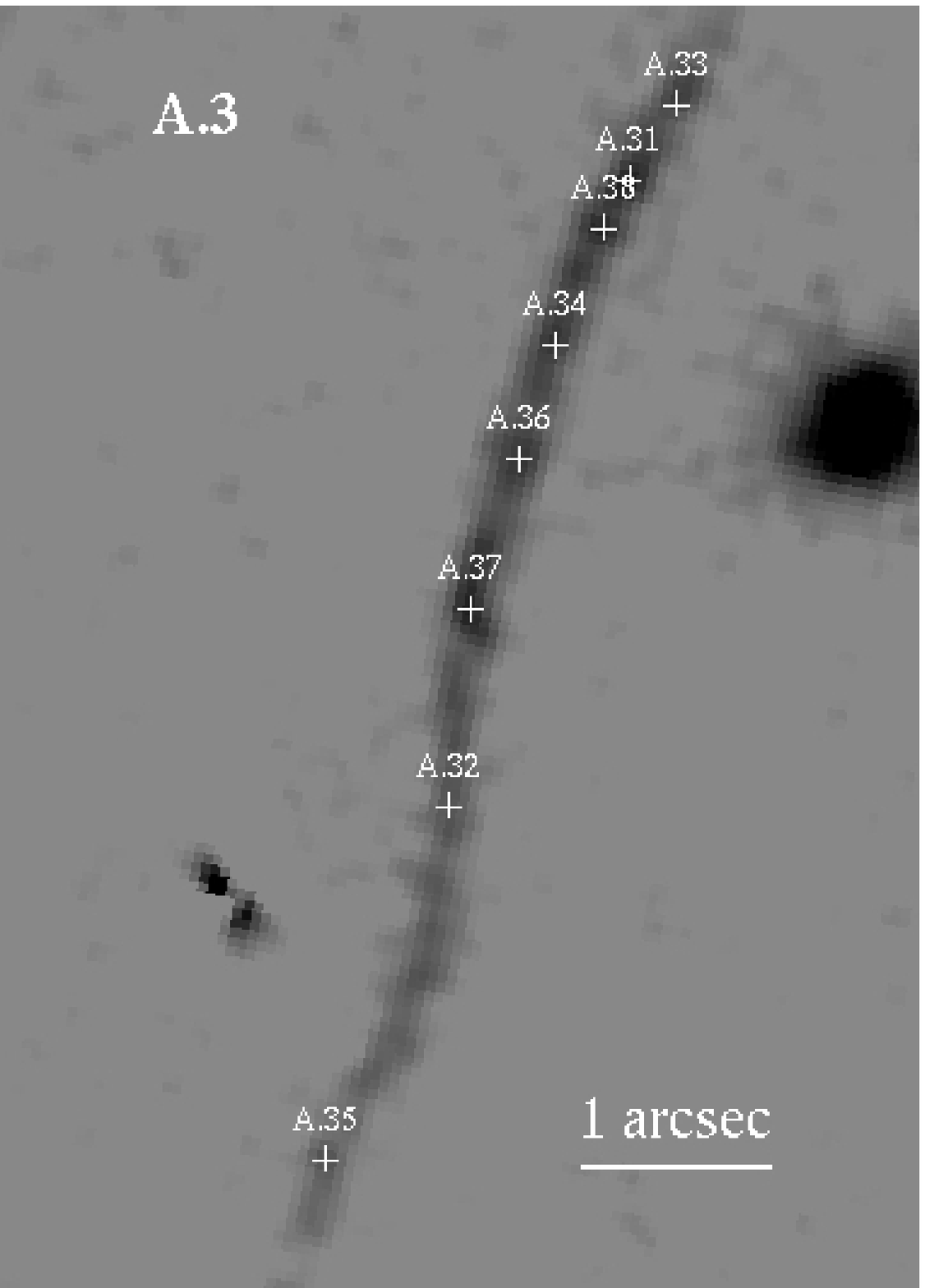}}
\vspace{0.2cm}
\hbox{\includegraphics[width=0.23\textwidth]{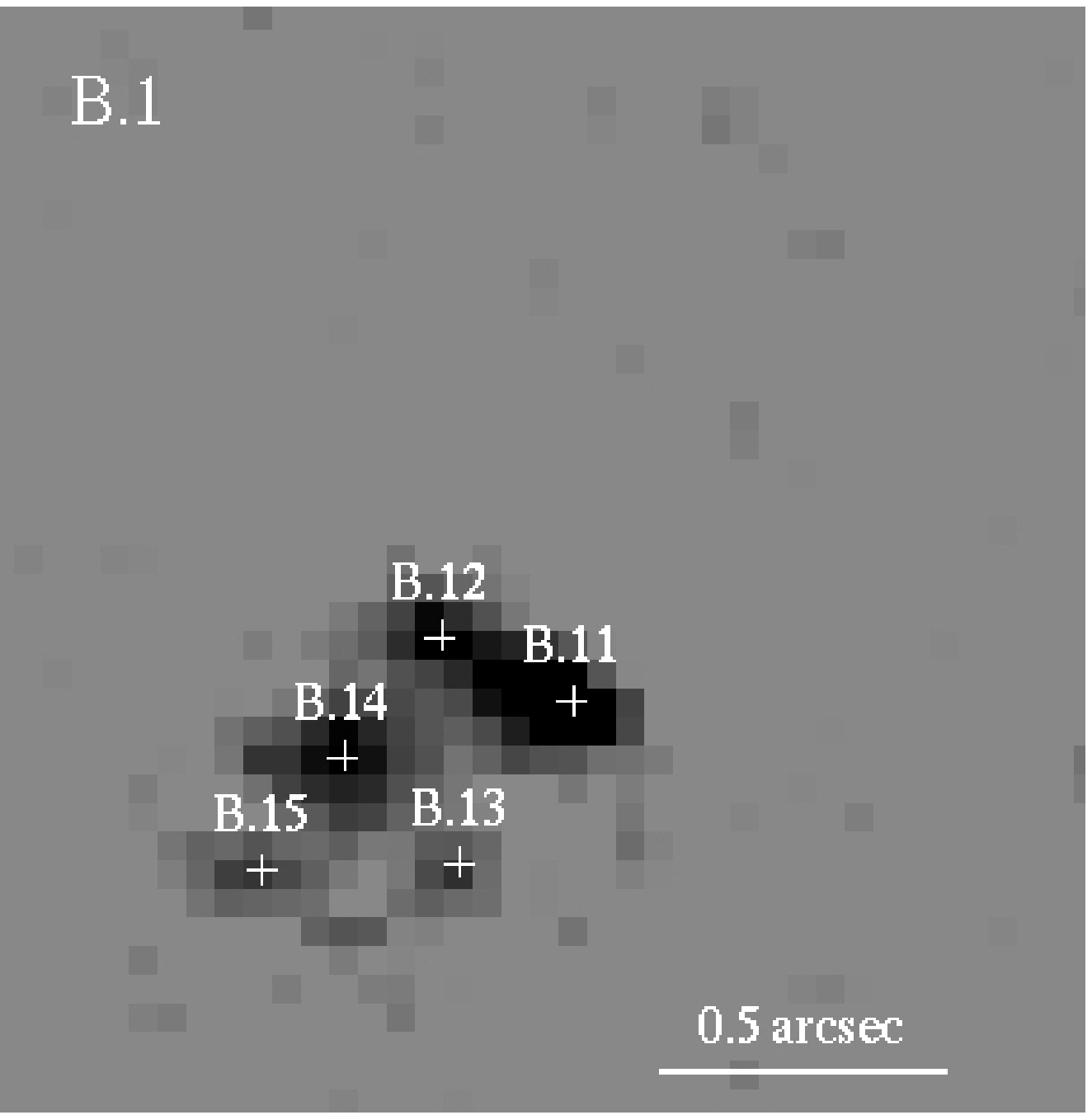}\hspace{0.18cm}\includegraphics[width=0.23\textwidth]{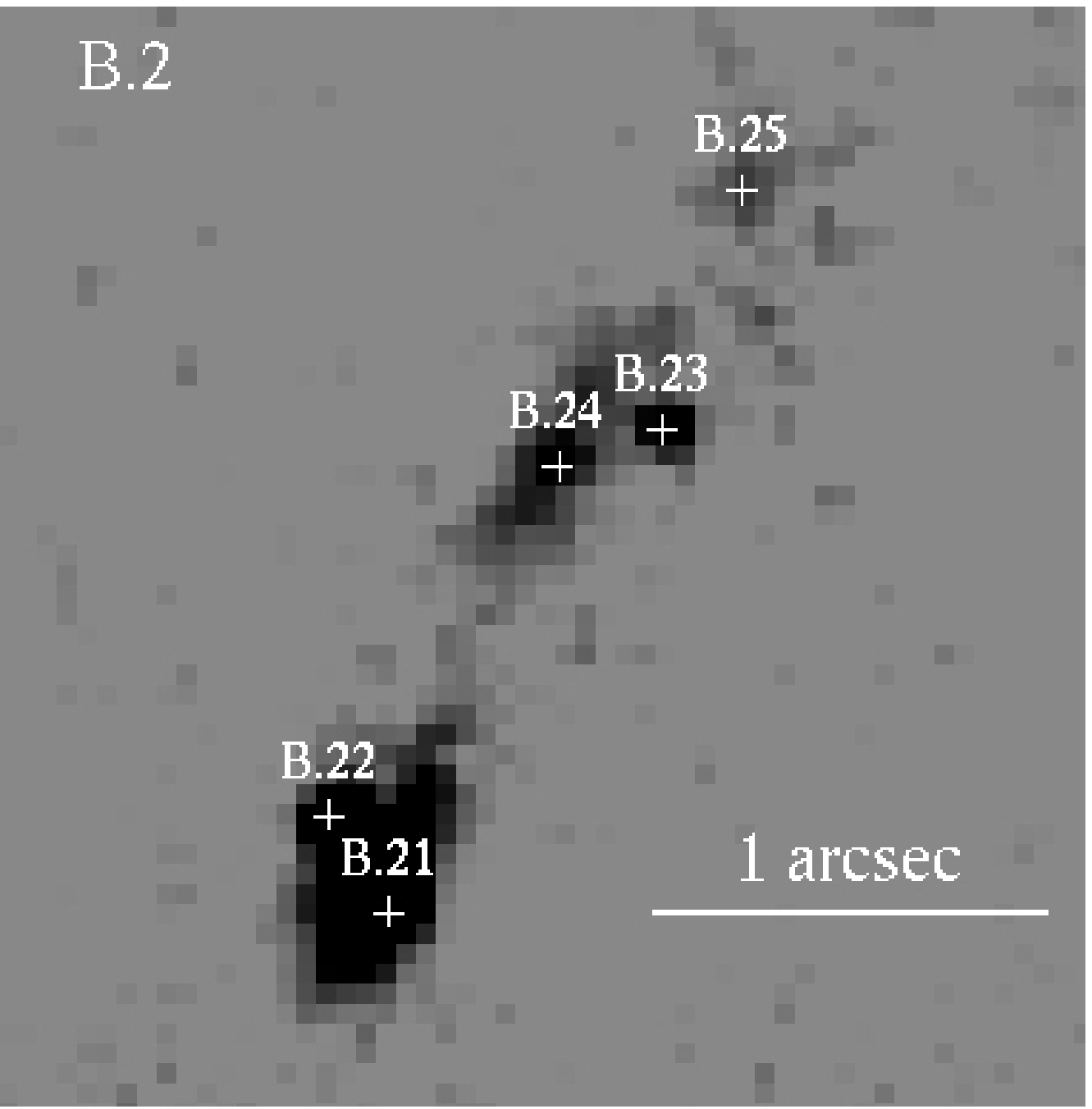}\hspace{0.18cm}\includegraphics[width=0.23\textwidth]{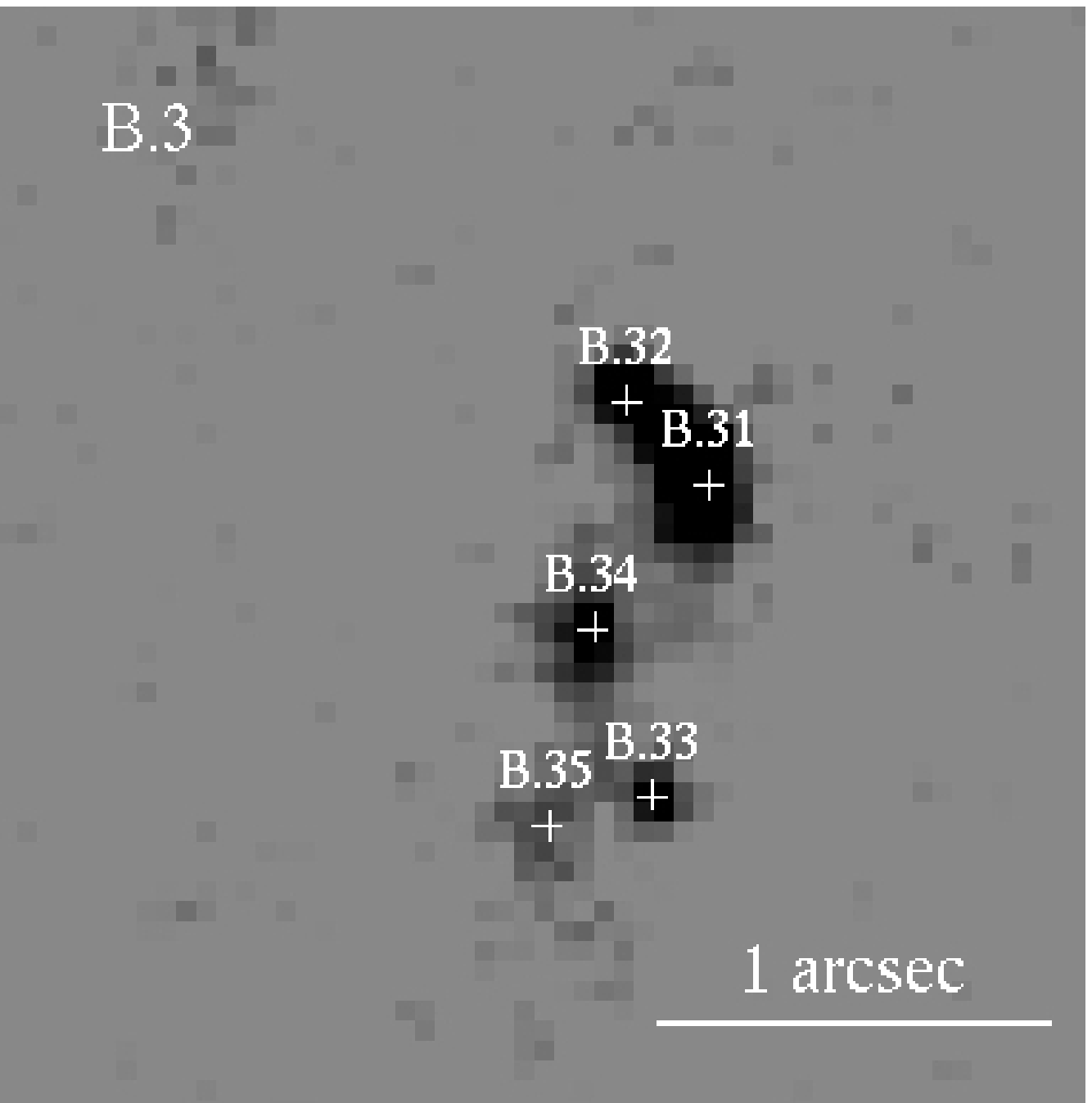}\hspace{0.18cm}\includegraphics[width=0.23\textwidth]{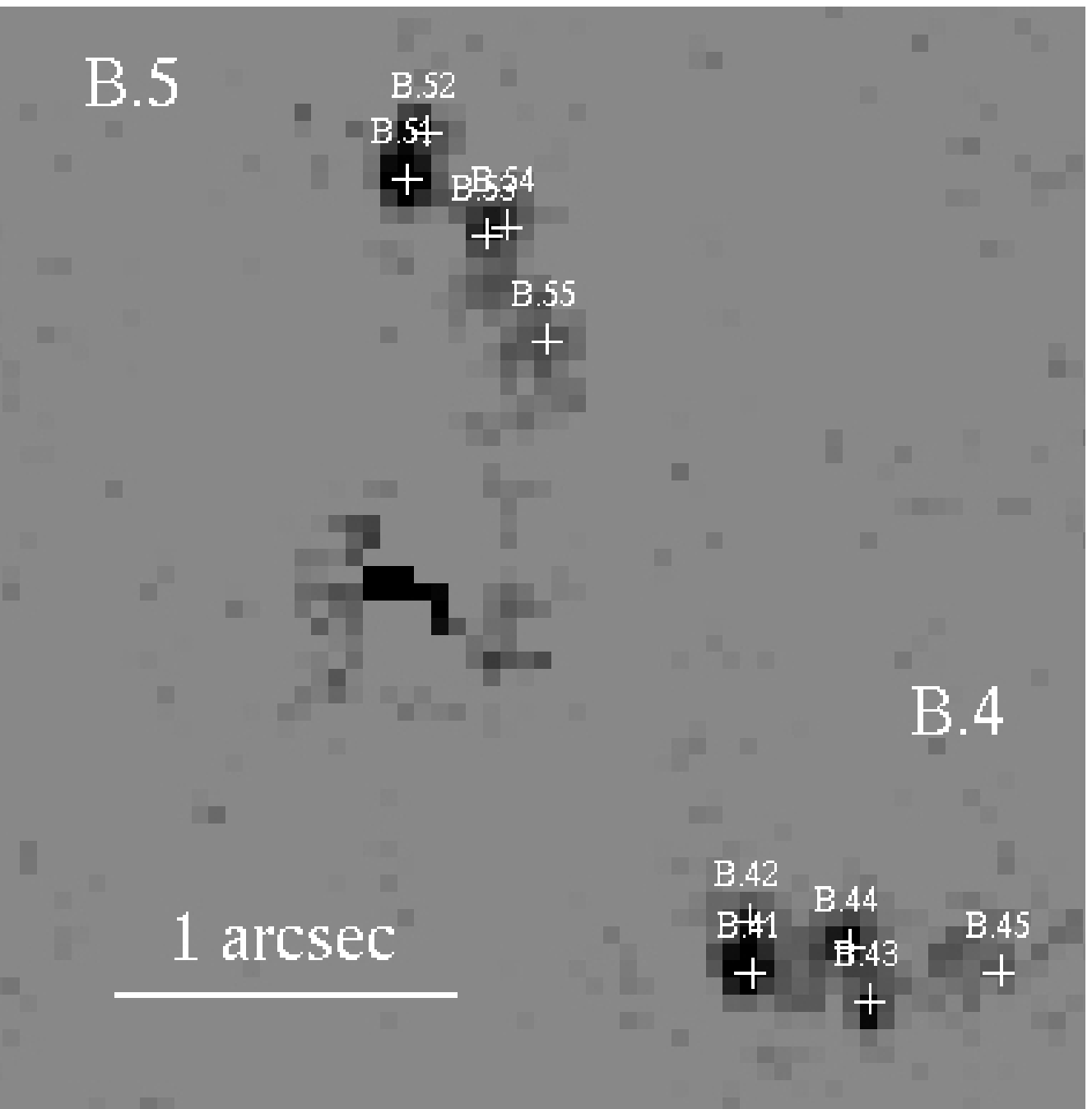}}
\end{center}
\caption[]{Details of the images used as constraints for the lens optimisation corresponding to the sources A and B (see Table~\ref{tab:input} in the Appendix). The first three panels on the left show the location of the knots selected along the tangential arc,  while the last four panels on the right show the  conjugate knots identified in the quintuplet system. All panels were selected in the galaxy-subtracted image, obtained through the two-dimensional fitting algorithm \textit{GALFIT} \citep{peng2002}. The images were smoothed with a 2 pixel FWHM Gaussian filter for a better visualisation.}
 \label{fig:snap}
\end{figure*}

We list in this appendix the coordinates of the input multiple image systems that we used to constrain the lens model for Abell~611. The position of the knots for the image systems corresponding to source A and B are shown in Fig.~\ref{fig:snap}.  
\begin{table*}[!hb]
\caption[]{Coordinates of the multiple images used as constraints in the lens model optimisation.}
\begin{center}
\begin{tabular*}{1.2\textwidth}{l@{\@{\hspace{1pt}}}l@{\@{\hspace{1pt}}}l@{\@{\hspace{3pt}}}l@{\@{\hspace{1pt}}}l@{\@{\hspace{1pt}}}l@{\@{\hspace{1pt}}}@{\@{\hspace{1pt}}}l@{\@{\hspace{1pt}}}l@{\@{\hspace{1pt}}}l@{\@{\hspace{1pt}}}l@{\@{\hspace{1pt}}}l@{\@{\hspace{1pt}}}l@{\@{\hspace{1pt}}}@{\@{\hspace{1pt}}}l@{\@{\hspace{1pt}}}@{\@{\hspace{1pt}}}l@{\@{\hspace{1pt}}}l@{\@{\hspace{1pt}}}}
\multicolumn{15}{c}{\textsc{Multiple image systems}} \\
\hline \hline
\multicolumn{15}{c}{System A}  \\
\textbf{A.1}  &   &    & 
\textbf{A.2}  &    &   & 
\textbf{A.3 } &    &   & 
 &  &  &
  & &  \\
\hline 
\textbf{1}  &  120.2372  &  36.061151  & 
\textbf{1}  &  120.24078  &  36.059618  & 
\textbf{1} &  120.24206  &  36.057857 & 
 &  &  &
  & &  \\
\textbf{2}  &  120.23744  &  36.060832  & 
\textbf{2} &  120.23927  &  36.060178  & 
\textbf{2}  &  120.24219  &  36.056926  \\
\textbf{3}  &  120.23715  &  36.061185  & 
\textbf{3 } &  120.24087  &  36.059567  & 
\textbf{3}  &  120.242  &  36.057975   &  & & &  & &\\ 
\textbf{4 } &  120.23727  &  36.06103  & 
\textbf{4}  &  120.24057  &  36.059648  & 
\textbf{4}  &  120.24214  &  36.057605   &  & & &  & &\\ 
\textbf{5}  &  120.23779  &  36.060597  & 
\textbf{5}  &  120.23838  &  36.060427  & 
\textbf{5 } &  120.2423  &  36.056396  &  & & &  & &\\ 
\textbf{6}  &  120.2373  &  36.060982  & 
\textbf{6}  &  120.24046  &  36.059611  & 
\textbf{6 } &  120.24217  &  36.057434   &  & & &  & &\\ 
\textbf{7}  &  120.23739  &  36.060874  & 
\textbf{7}  &  120.23999  &  36.05979  & 
\textbf{7}  &  120.24221  &  36.05721   &  & & &  & &\\ 
\textbf{8}  &  120.23716  &  36.061085  & 
\textbf{8}  &  120.24068  &  36.059646  & 
\textbf{8 } &  120.24209  &  36.057782   &  & & &  & &\\ 
\hline
 \multicolumn{15}{c}{System B} \\
 \textbf{B.1}  &   &    & 
\textbf{B.2 } &    &   & 
\textbf{B.3}  &    &   & 
\textbf{B.4}  &   &   & 
\textbf{B.5}  &   &   \\ 
 \hline

\textbf{1}  &  120.24114  &  36.05811  & 
\textbf{1 } &  120.24183  &  36.055047  & 
\textbf{1}  &  120.23566  &  36.054089  & 
\textbf{1}  &  120.2323  &  36.061433  & 
\textbf{1}  &  120.236  &  36.054733 \\ 
\textbf{2 } &  120.24119  &  36.058177  & 
\textbf{2 } &  120.2419  &  36.055106  & 
\textbf{2}  &  120.23566  &  36.054132  & 
\textbf{2 } &  120.23238  &  36.061464  & 
\textbf{2 } &  120.23598  &  36.054772  \\ 
\textbf{3 } &  120.2409  &  36.058445  & 
\textbf{3}  &  120.24188  &  36.05483  & 
\textbf{3}  &  120.23554  &  36.054066  & 
\textbf{3}  &  120.23237  &  36.061354  & 
\textbf{3}  &  120.23592  &  36.054688  \\
\textbf{4 } &  120.24099  &  36.05842  & 
\textbf{4 } &  120.24193  &  36.054947  & 
\textbf{4 } &  120.23556  &  36.054111  & 
\textbf{4 } &  120.23244  &  36.061406  & 
\textbf{4}  &  120.2359  &  36.054695  \\
\textbf{5 } &  120.24083  &  36.058612  & 
\textbf{5}  &  120.24197  &  36.054809  & 
\textbf{5}  &  120.23541  &  36.054089  & 
\textbf{5}  &  120.23249  &  36.061351  & 
\textbf{5}  &  120.23586  &  36.054602  \\
\hline
 \multicolumn{15}{c}{System C}\\
 \textbf{C.1 } &   &    & 
\textbf{C.2 } &    &   & 
 & &   &
   & & &
     & & \\

 \hline
\textbf{1 } &  120.24193  &  36.056063  & 
\textbf{1 } &  120.23206  &  36.061931 & 
 & &   &
   & & &
     & & \\
     \hline
 \multicolumn{15}{c}{System D$^\star$}\\
 \textbf{D.1} && & \textbf{D.2} && & \textbf{D.3} &
 & & \textbf{D.4} && & & &  \\
 \hline
\textbf{1} &120.23551 & 36.060743 & \textbf{1} &120.2376 & 36.060501 & \textbf{1} &120.24321 & 36.053464 & \textbf{1} &120.23417 & 36.055649 & & &  \\
\textbf{2} &120.23748 & 36.060537 &\textbf{2} &120.2357 & 36.060719 & &   &
   & & &
     & & \\
\textbf{3}&120.23549 & 36.060665 & \textbf{3} &120.23746 & 36.06045 & &   &
   & & &
     & & \\
\textbf{4}&120.23561 & 36.060677 & \textbf{4} &120.23742 & 36.060507 &\textbf{4}&120.24326 & 36.053421 &\textbf{4}&120.23409 & 36.055685 & & &  \\

\hline \hline
\end{tabular*}

\end{center}
\label{tab:input}	     
\tablefoot{Coordinates are in degrees WCS. The redshifts of the sources 
corresponding to System A, System B and System C are $ z_{\rm A}=0.908\pm 0.005$, 
$z_{\rm B}=2.06\pm0.02$  and  $ z_{\rm C}=2.59\pm0.01$, respectively \citep{richard2009}. \\
$^\star$ System D is not yet confirmed spectroscopically: for this reason, 
we included it as constraint only in one dedicated test. 
The redshift estimates for the knot groups of this system are listed in Table~\ref{tab:tab_l1}.}
\end{table*}

\clearpage
\section{Comparison between predicted image configurations}
 \label{app:comp_im}
 
 In this appendix we compare the observed strong lensing system of Abell~611 to the lensed  images predicted by our best-fit models. The top left panel shows the \textit{ACS/HST} image of Abell~611 (filter F606W), in which the galaxies were subtracted through the software \textit{Galfit}. The last seven panels show the images predicted by our best-fitting  lens models,  corresponding to the cases listed in Table~\ref{tab:tab_l1}.  
\begin{figure}[hb]
\begin{center}
\hbox{\includegraphics[width=0.32\textwidth]{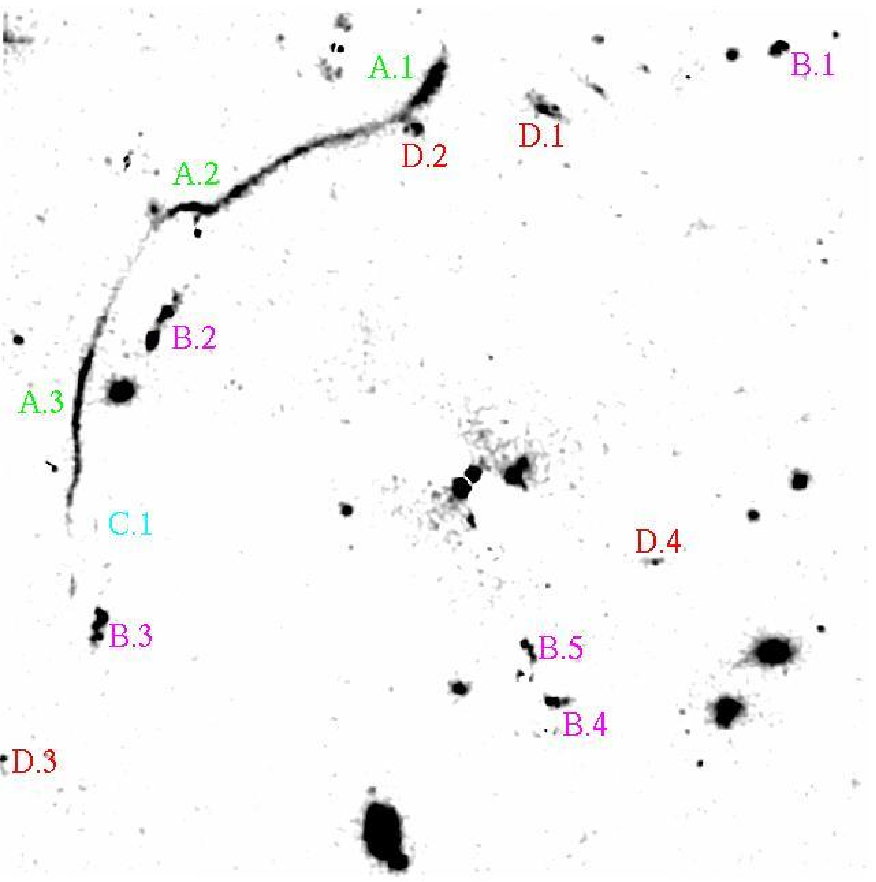} \includegraphics[width=0.32\textwidth]{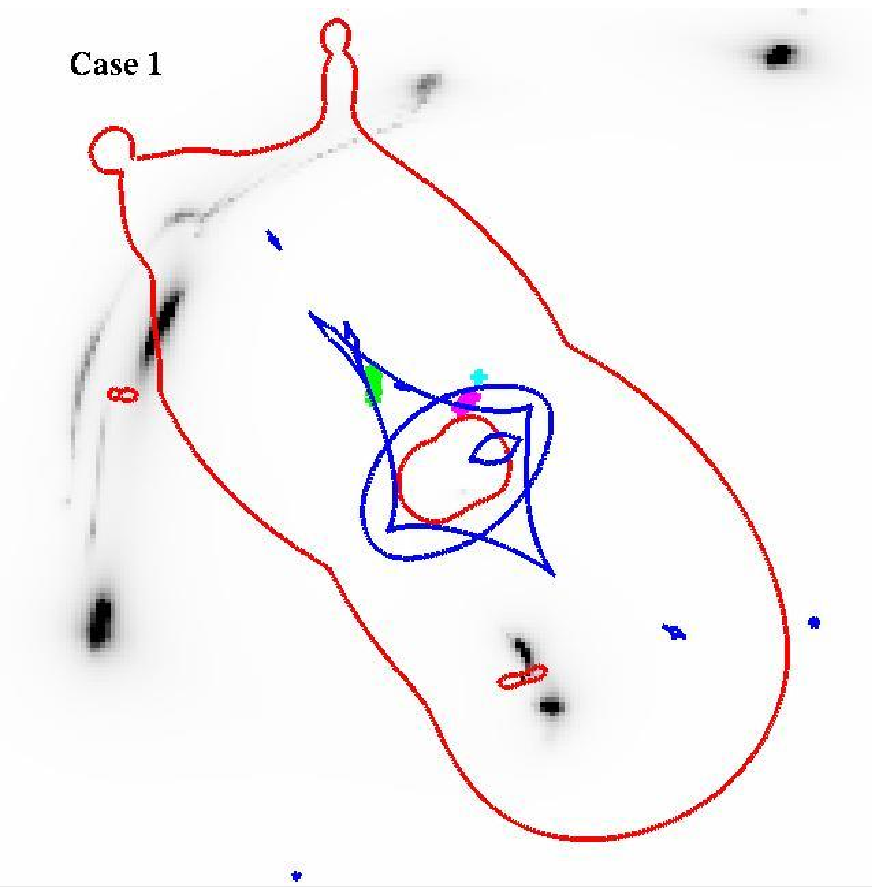}\includegraphics[width=0.32\textwidth]{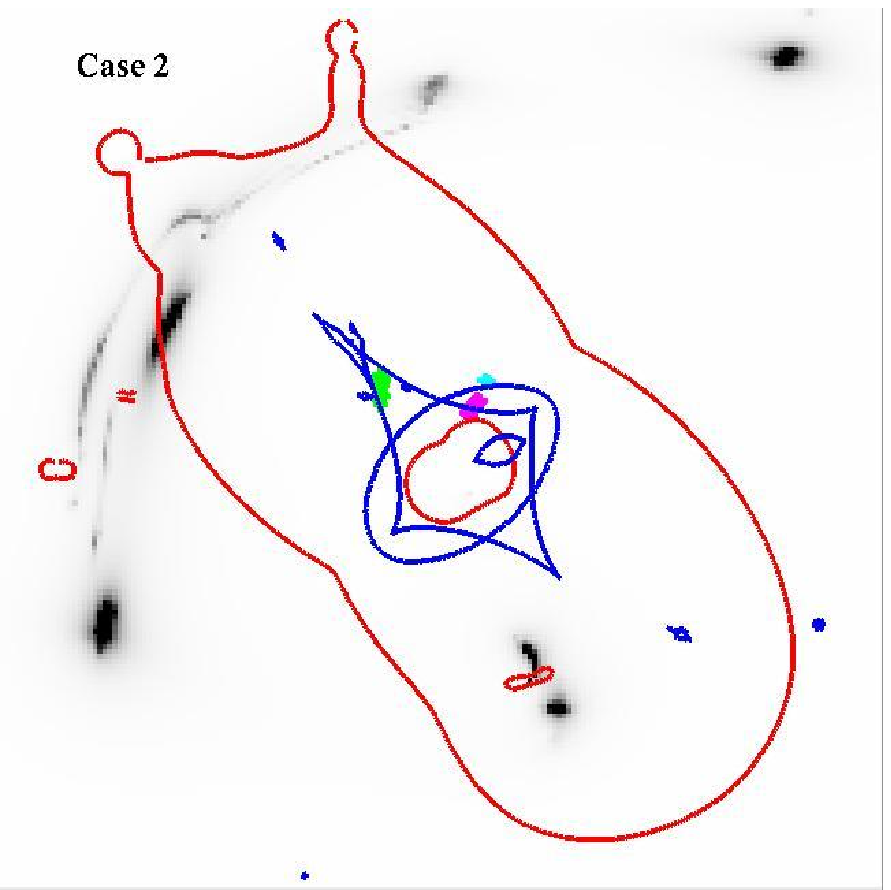}} 
\hbox{\includegraphics[width=0.32\textwidth]{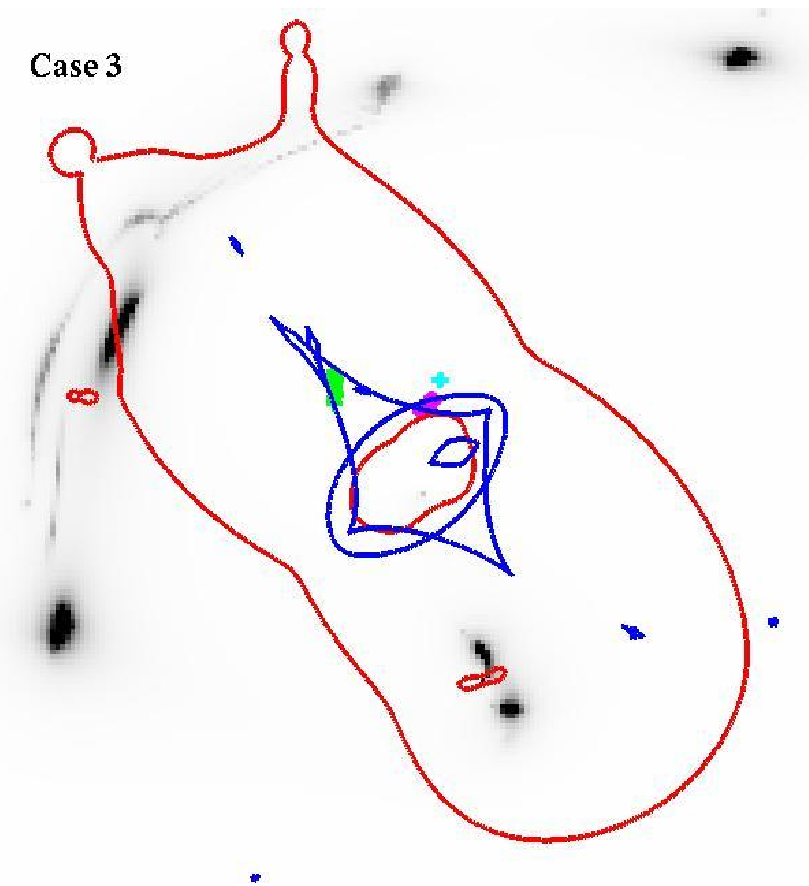}\includegraphics[width=0.32\textwidth]{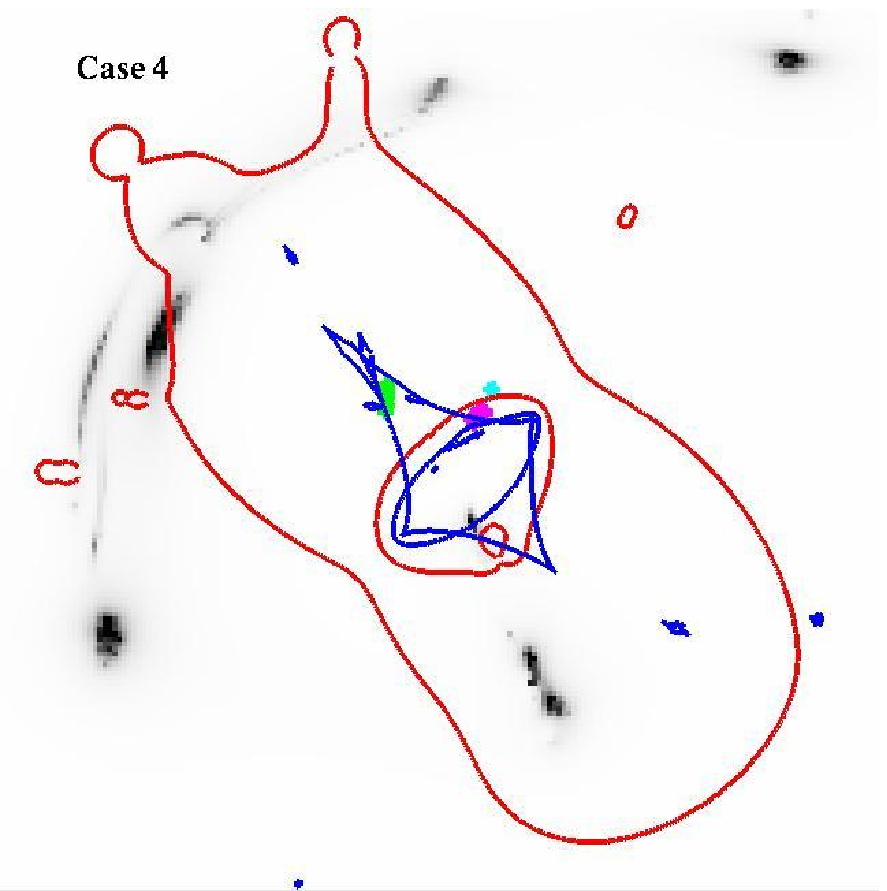}\includegraphics[width=0.32\textwidth]{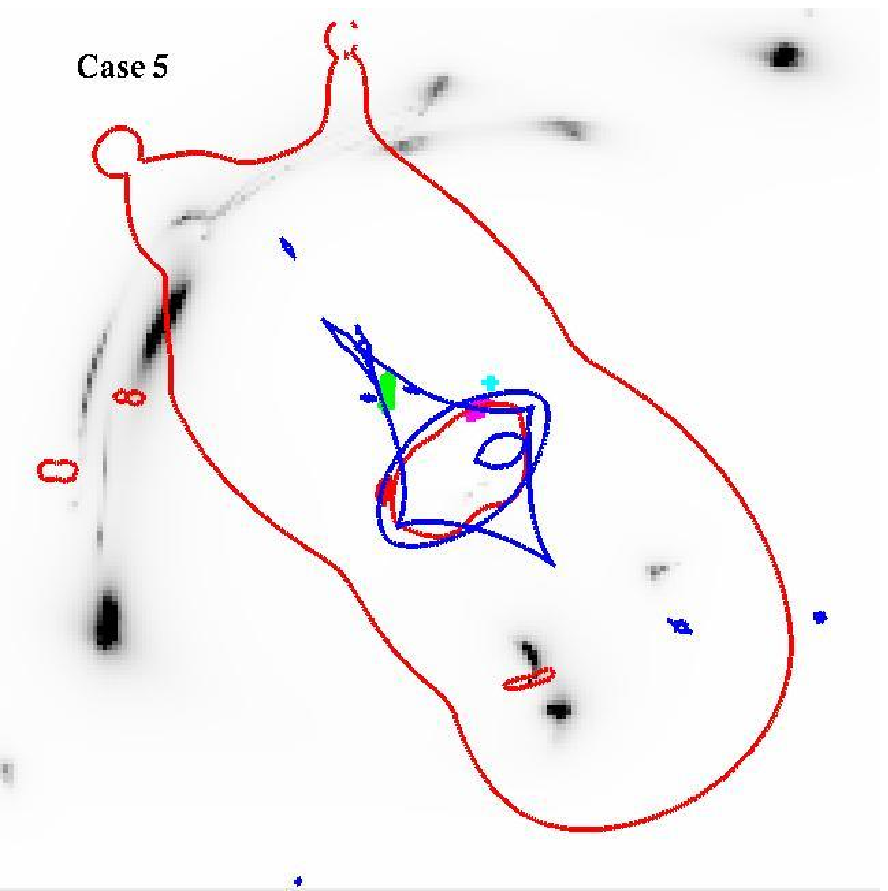}}
\hbox{\hspace{3.5cm}\includegraphics[width=0.32\textwidth]{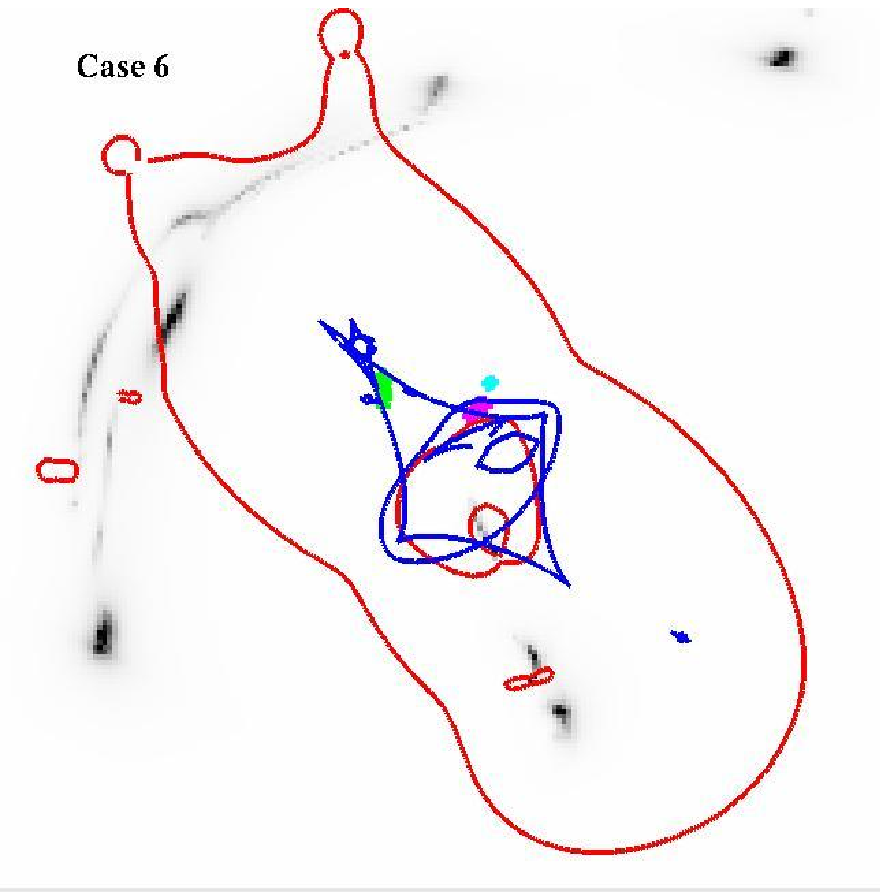}
 \includegraphics[width=0.32\textwidth]{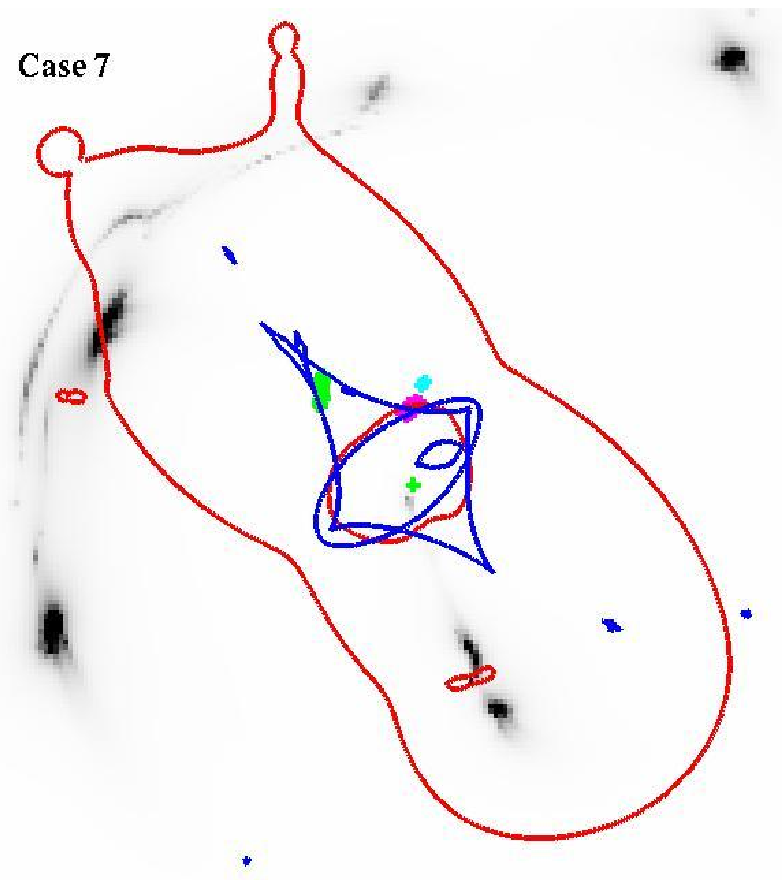}}
\end{center}
\caption[]{Comparison between the images predicted by our best-fit models and the observed lensed features. \textit{[Top left panel]} Galaxy-subtracted \textit{ACS/HST} image of Abell~611. The identified (A,B,C) and candidate (D) lensed systems are marked in green, magenta, cyan, and red.  \textit{[Top middle, top right, middle and bottom panels]} The predicted critical (caustic) lines  are overlaid in red (blue) on the images predicted by the best-fit models  listed in Table~\ref{tab:tab_l1}. They refer to a source plane at  redshift  $z_A=0.908$.   The  predicted sources are drawn in the last seven panels with the same colour as the corresponding image system  in the top left panel. The conjugated knots used as constraints  are  shown in Fig.~\ref{fig:sysgen}, whereas their coordinates are listed in Table~\ref{tab:input}.  The field dimensions of the  panels are $\simeq [35.2 \times 35.1]$ arcsec; all fields are WCS-aligned. The flux scale is the same in the last seven panels. 
}
 \label{fig:resl}
\end{figure}

\end{document}